\documentclass[aps,prd,reprint,onecolumn,groupedaddress,longbibliography,notitlepage]{revtex4-1}
\usepackage[utf8]{inputenc}
\usepackage[T1]{fontenc}

\newcommand{\Rey}{Re}

\usepackage{siunitx} 
\DeclareSIUnit[]{\pixel}{px}

\usepackage{placeins}

\usepackage{graphicx}
\usepackage{epstopdf, epsfig}
\usepackage{latexsym}
\usepackage{amssymb}
\usepackage{mathtools}
\usepackage[outline]{contour}
\newcommand{\graysquare}{\contour{black}{$\color[rgb]{0.5,0.5,0.5}{\blacksquare} \,\,$}}

\newcommand{\blackcircle}{\contour{black}{$\color[rgb]{0,0,0}{\bullet} \,\,$}}

\newcommand{\whitecircle}{\contour{black}{$\color[rgb]{1,1,1}{\bullet} \,\,$}}

\newcommand{\graytriangleup}{\contour{black}{$\color[rgb]{0.8,0.8,0.8}{\blacktriangle} \,$}}

\usepackage{tikz}
\newcommand\solidrule{\protect\tikz[baseline]{\protect\draw[line width=0.7mm] (0,.5ex)--++(1,0) ;}}
\newcommand\finesolidrule{\protect\tikz[baseline]{\protect\draw[line width=0.4mm] (0,.5ex)--++(1,0) ;}}
\newcommand\dashedrule{\protect\tikz[baseline]{\protect\draw[line width=0.7mm,dashed] (0,.5ex)--++(1,0) ;}}
\newcommand\finedashedrule{\protect\tikz[baseline]{\protect\draw[line width=0.4mm,dashed] (0,.5ex)--++(1,0) ;}}
\newcommand\dashdotrule{\protect\tikz[baseline]{\protect\draw[line width=0.7mm,dash dot] (0,.5ex)--++(1,0) ;}}

\begin{document}

\title{A mass entrainment-based model for separating/reattaching flows}

\author{F. Stella}
\author{N. Mazellier}
\email{nicolas.mazellier@univ-orleans.fr}
\author{P. Joseph}
\altaffiliation{Presently at Univ. Lille, CNRS, ONERA, Centrale Lille, Arts et M\'{e}tiers Paris Tech, FRE 2017, Laboratoire de M\'{e}canique des Fluides de Lille - Kamp\'{e} de F\'{e}riet, Lille, F59000, France}
\author{A. Kourta}

\affiliation{University of Orl\'{e}ans, INSA-CVL, PRISME, EA 4229, F45072, Orl\'{e}ans, France}

\date{\today}

\begin{abstract}
Recent studies have shown that entrainment effectively describes the behaviour of natural and forced separating/reattaching flows developing behind bluff bodies, potentially paving the way to new, scalable separation control strategies. In this perspective, we propose a new interpretative framework for separated flows, based on mass entrainment. The cornerstone of the approach is an original model of the mean flow, representing it as a stationary vortex scaling with the mean recirculation length. We test our model on a set of mean separated topologies, obtained by forcing the flow over a descending ramp with a rack of synthetic jets. Our results show that both the circulation of the vortex and its characteristic size scale simply with the intensity of the backflow (i.e. the amount of mass going through the recirculation region). This suggests that the vortex model captures the essential functioning of mean mass entrainment, and that it could be used to model and/or predict the mean properties of separated flows. In addition, we use the vortex model to show that the backflow (i.e. an integral quantity) can be estimated from a single wall-pressure measurement (i.e. a pointwise quantity). Since the backflow also appears to be anticorrelated to the characteristic velocity of the synthetic jets, this finding suggests that industrially deployable, closed-loop control systems based on mass entrainment might be within reach.  
\end{abstract}

\maketitle
\section{Introduction}\label{sec:introVortex}
Separating/reattaching flows are one important source of aerodynamic losses in many industrial flows, one common example being the large shape drag of bluff bodies such as long-haul, heavy ground vehicles \cite{seifert2015}. 
In this respect, understanding and controlling these flows is of primary importance for improving performances of industrial systems, in particular in the present context of increasingly stringent environmental regulations. 
Prototypical separating/reattaching flows developing on simple geometries, such as the backward facing step (BFS) or ramps of various shapes, have been studied extensively for several decades, giving us a relatively complete general understanding of their functionning. 
The fully turbulent separating/reattaching flow generally presents one large recirculation region \cite{armaly1983}, which extends from the separation point (usually fixed at the salient edge of, say, the BFS) to the reattachment point.
A separated shear layer develops between the recirculation region and the free flow, growing in thickness until it hits the wall at the reattachment point. Over the first half of the recirculation region, the separated shear layer seems to behave almost as a free shear layer \cite{simpson1989}. Similarities include the value of its growth rate \cite{dandois07,stella2017} as well as the presence of a convective instability (\cite{debien14,kourta15}, among others) reminding the convection of large-scale structures reported by \citet{brownRoshko74} in free shear layers. 
Downstream of the reattachment point, the flow slowly relaxes to a new boundary layer \cite{le1997}.

One primary approach to mitigate aerodynamic losses caused by separating/reattaching flows has focused on artificially modifying their induced pressure distribution, for example in order to reduce drag \cite{barros2016}. This often comes down to controlling the shape of the recirculation region, its size, or both. Usually, the mean reattachment length $L_R$, defined as the streamwise distance between the mean separation point and the mean reattachment point, is considered an appropriate, synthetic indicator of these geometric properties of the recirculation region. 
Many techniques have been proposed to modify $L_R$, ranging from passive devices as vortex generators \cite{pujals2010}, to active systems such as steady suction \cite{roumeas2009} or blowing \cite{donovan1997}, pulsed jets \cite{joseph2012}, plasma actuators \cite{thomas2008} and synthetic jets \cite{kourta2013}. Among these methods, those based on a periodic forcing have received particular attention, because of their ability to interact with the instabilities of the separated shear layer. 
A large corpus of experimental works \cite{shimizu1993,sigurdson1995,chun1996,glezer2005,parezanovic2015} as well as numerical studies \cite{dandois07} has shown that periodic actuators can be more or less effective at modifying the shape of the separation, depending on how the frequency of the forcing compares to the characteristic frequency of the natural convective instability. With respect to this natural threshold, low actuation frequencies generally create a train of coherent counterotating vortices \cite{berk2017}, which enhances the growth of the separated shear layer and reduces $L_R$ \cite{adamsJohnstonPart1}. High actuation frequencies, instead, tends to dampen shear layer instabilities, thus hindering its growth and increasing $L_R$.

Despite these promising results, it is an empirical fact that industrially operative, active flow control systems are rare.
In particular, one critical issue that is yet to be solved concerns their scalability. Indeed, the scaling parameters of the controlled flow are of fundamental importance, even with black-box approaches, to guarantee reliable deployment to full-scale applications, but they are still poorly understood. Quite remarkably, even in the case of \textit{unperturbed} separating/reattaching flows we miss a clear understanding of the scaling laws of even simple, mean-field features such as $L_R$.
Indeed, although some $L_R$ patterns have been identified which seem consistent across experiments \cite{armaly1983,nadge14}, their practical use is often limited, because the behaviour of $L_R$ appears to be significantly influenced by complex interactions of a large number of factors such as geometry \cite{durst1981,ruck1993}, free flow turbulence \cite{adamsJohnstonPart2}, and the ratio between the thickness of the incoming boundary layer and the height of the step \cite{adamsJohnstonPart1}.

In this respect, recent works suggest that simpler descriptions of the behaviour of both unperturbed and controlled separating/reattaching flows might be obtained by considering the role of mass or momentum entrainment. 
Picking up from the theory of \citet{chapman1958}, \citet{stella2017} analysed the exchanges of mass within the separated flow behind a descending ramp, for several different values of the parameter $\Rey_\theta = U_\infty \theta/\nu$, where $U_\infty$ indicates free-stream velocity, $\theta$ is the momentum thickness of the incoming boundary layer and $\nu$ is the kinematic viscosity of the fluid. 
They found that $L_R^* = L_R/h$, with $h$ being the height of the ramp, scales as $\Rey_\theta^m$, where $m$ depends on the turbulent state of the incoming boundary layer. More interestingly,  $L_R^*$ is approximately linearly proportional to the normalised backflow $\dot{m}_R^* = \dot{m}_R/\left(\rho U_\infty h\right)$, $\rho$ being density, that is the flux of mass that goes through the recirculation region.
In other words, \citet{stella2017} show that the characteristic length scale of separating/reattaching flows scales simply with the backflow $\dot{m}_R^*$, while the relatively complex dependency on $\Rey_\theta$ can be confined to the behaviour of $\dot{m}_R^*$.
Results reported by \citet{berk2017} convey a similar idea. These researchers used synthetic jets to control $L_R^*$ in a BFS flow. To compare the action of the jet for four different values of actuation frequency, they measured the amount of mean field momentum entrained from the free flow in a large control volume, encompassing the entire recirculation region, by the train of vortices generated by the jet. Interestingly, they found that the evolution of $L_R^*$ becomes linear when it is expressed in function of this momentum entrainment, regardless to how the actuation frequency compares to the natural convective instability of the flow. In addition, \citet{berk2017} used phased-locked PIV to analyse the convected vortices, observing that the amount of entrained momentum depends on the frequency-driven nature of the interactions between successive vortices. Once again, entrainment provides a relatively simple description of the behaviour of $L_R^*$, independently of how it is itself affected by the variable controlling the flow.

Observations reported by \citet{berk2017} and \citet{stella2017} on the variation of $L_R^*$ contribute to draw attention on the importance of entrainment in the functionning of separating/reattaching flows. However, they do not define an entrainment-based, coherent interpretative framework for these flows.
This work aims at filling this gap, by proposing a new model of separating/reattaching flows that puts mass entrainment at the center of the picture.
In this respect, analysis of mass entrainment is privileged for its simplicity, but also because verifying continuity seems a more pertinent approach to the study of the behaviour of $L_R^*$ (i.e. the characteristic size of a \textit{closed region}) than the investigation of mean momentum transfer.
We stress that, for developing and discussing our model, we will exclusively focus on the \textit{mean flow}, intended in the sense of the Reynolds Averaged Navier-Stokes (RANS) equations. Accordingly, throughout this work we will conform to the so-called Reynolds decomposition of the velocity field, and to its standard notation. For example, the instantaneous wall-normal velocity component $v$ will be written as:
\begin{equation}
v = V + v^\prime,
\end{equation}
where $V$ and $v^\prime$ are the mean and fluctuating wall-normal velocities, respectively. The same decomposition and notation apply to the streamwise velocity component $u$. In the remainder of the paper, the symbol $^*$ will be used to indicate normalisation on $U_\infty$, or $h$, or both, depending on the dimensions of the normalised quantity. 

\begin{figure}
  \centering
  \includegraphics[width=\linewidth]{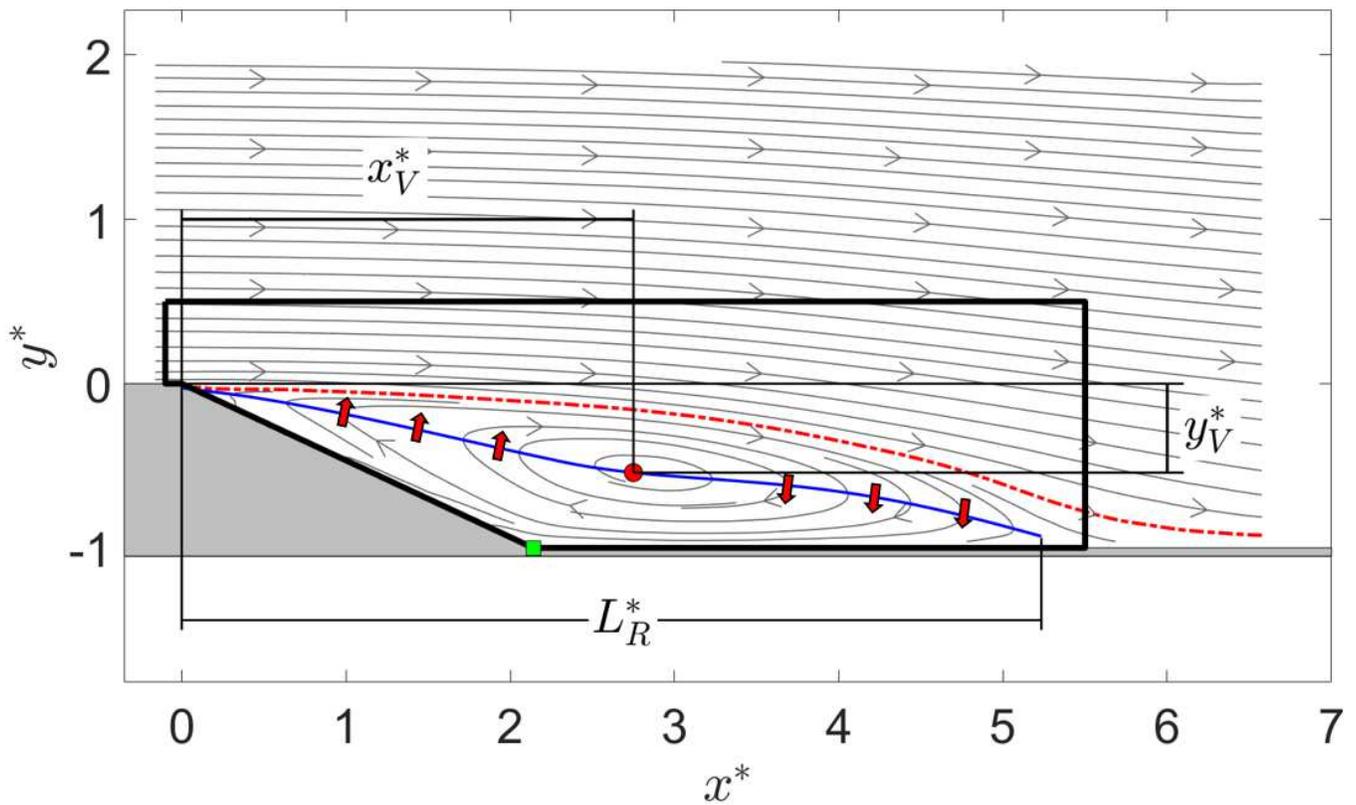} 
\caption{Streamlines of the baseline separated flow. Symbols: \protect{\color[rgb]{0.4,0.4,0.4}{\solidrule}}$\,\,$(gray online) streamlines; \protect\solidrule$\,\,$(black online) closed contour $S_V$, used for the computation of $\Gamma^*$; \protect{\color[rgb]{0,0,1}{\solidrule}} (blue online) mean separation line; \protect{\color[rgb]{1,0,0}{\dashdotrule}}$\,\,$(red online) separating streamline; \protect\contour{black}{$\color[rgb]{1,0,0}{\bullet} \,\,$} center of the vortex, identified as the point of the mean separation line where $V=0$; \protect\contour{black}{$\color[rgb]{0,1,0}{\blacksquare} \,\,$} position of the pressure tap measuring $P_b$ (see Sec. \ref{sec:backflowObs}). Red arrows represent mass fluxes through the mean separation line.}
\label{fig:vortex}
\end{figure} 

Let us consider Figure \ref{fig:vortex}, which shows the streamline representation of the mean flow investigated in this study. This is a massive turbulent separation, developing behind a descending ramp geometrically similar to the one used in \citet{stella2017}. In contrast with that study, however, in the present experiment $L_R^*$ can be forced with a rack of synthetic jets, as done in \citet{berk2017}. 
In Figure \ref{fig:vortex}, streamline patterns can be interpreted as those of a large spanwise vortex. Such vortex appears as the dominant feature of the mean separated flow. Its elongated shape, fitting between the wall and the dividing streamline (\cite{chapman1958} among others), is approximately divided in half by the mean separation line \cite{eatonJohnson1981}. 
Then, the lower part of the vortex covers the entire mean recirculation region, while its upper one corresponds to a sizeable portion of the mean separated shear layer.
This being so, $L_R^*$ is the characteristic streamwise length of the vortex.
In addition, streamlines indicate that the vortex rotates clockwise. Then, the vortex \textit{entrains} fluid from the recirculation region to the shear layer in a neighbourhood of the mean separation, and viceversa in a neighbourhood of the mean reattachment point (represented with red arrows in Figure \ref{fig:vortex}).
This is exactly the general description of the backflow provided by \citet{chapman1958} and \citet{stella2017}, so that $\dot{m}_R^*$ seems to be related to the amount of mass put in rotation by the vortex.
All in all, the spanwise vortex seems to be representative of the main topological features of the mean flow, in particular $L_R^*$, and to coherently include $\dot{m}_R^*$ in the picture. 
As so, we argue that modeling mean separating/reattaching flows with a stationary vortex might be the cornerstone of the new, (mass) entrainment-based investigation framework that is anticipated by \citet{berk2017} and \citet{stella2017}.

The first objective of this paper is to lay the fundation of such \textit{vortex model}. In its simplest forms, the velocity field induced by a vortex can be determined from its circulation $\Gamma_V^*$ and its characteristic size $L_R^*$ \cite{batchelor2000}. For the vortex to support the analysis of mean separating/reattaching flow in terms of mass entrainment, it is then key to relate both $\Gamma_V^*$ and $L_R^*$ to the backflow. To address this issue, we investigate the relations connecting these quantities on a set of different mean separated flow topologies, obtained by forcing the baseline ramp flow of Figure \ref{fig:vortex} with the synthetic jets. 
In addition to characterising the vortex model, our hope is to confirm the linear $L_R^*$ trend found by \citet{berk2017}, in their analysis of momentum entrainment from the free flow. This would reduce results reported by \citet{berk2017} and by \citet{stella2017} to a single entrainment description, based on $\dot{m}_R^*$ and supported by the vortex model of the flow.

An important problem affecting any entrainment-based approach to the control of separating/reattaching flows concerns the observability of entrainment.
Indeed, directly measuring $\dot{m}_R^*$ (as well as any other mass or momentum flux) requires to reconstruct large portions of at least a bidimensional mean velocity field. 
This is unfeasible in most industrial applications, in which flow sensing can usually rely only on sparse, pointwise input, one typical example being wall-pressure information sensed by flush-mounted pressure taps. 
In spite of the observed linear $L_R^*$ trends, then, the reconstruction of $\dot{m}_R^*$ appears to be a real showstopper for any practical application of an entrainment-based approach to flow control.
Anyway, the vortex model might offer new ideas to tackle this problem, since the pressure distribution induced by a vortex appears to be, at least to a certain extent, related to its circulation and its topology \cite{hunt1988,jeong1995}. Then, the second objective of this paper is to use the vortex model to propose a correlation between $\dot{m}_R^*$ and wall-pressure, that might serve to develop simple, industrially deployable estimators of $\dot{m}_R^*$. 

The paper is organised as follows. Section \ref{sec:expSetup} describes the experimental set-up, including the descending ramp and the baseline flow. The characteristics of the synthetic jets as well as their effects on the scaling of forced flows are discussed in section \ref{sec:actuDev}. Section \ref{sec:LR_backflow} is dedicated to the backflow: it covers the computation of $\dot{m}_R^*$ and its correlation to both $\Gamma_V^*$ and $L_R^*$. The reconstruction of $\dot{m}_R^*$ from wall-pressure measurements is discussed in section \ref{sec:backflowObs} and conclusions and perspectives are given in section \ref{sec:concluVortex}. In addition, some annexed aspects of mass entrainment are presented in Appendix.

\section{Experimental set-up}\label{sec:expSetup}
This section describes the experimental model, the baseline flow and the measuring devices used in this study.
\subsection{Model, wind tunnel and baseline flow}
\begin{figure}
  \centering
  \includegraphics[width=\linewidth]{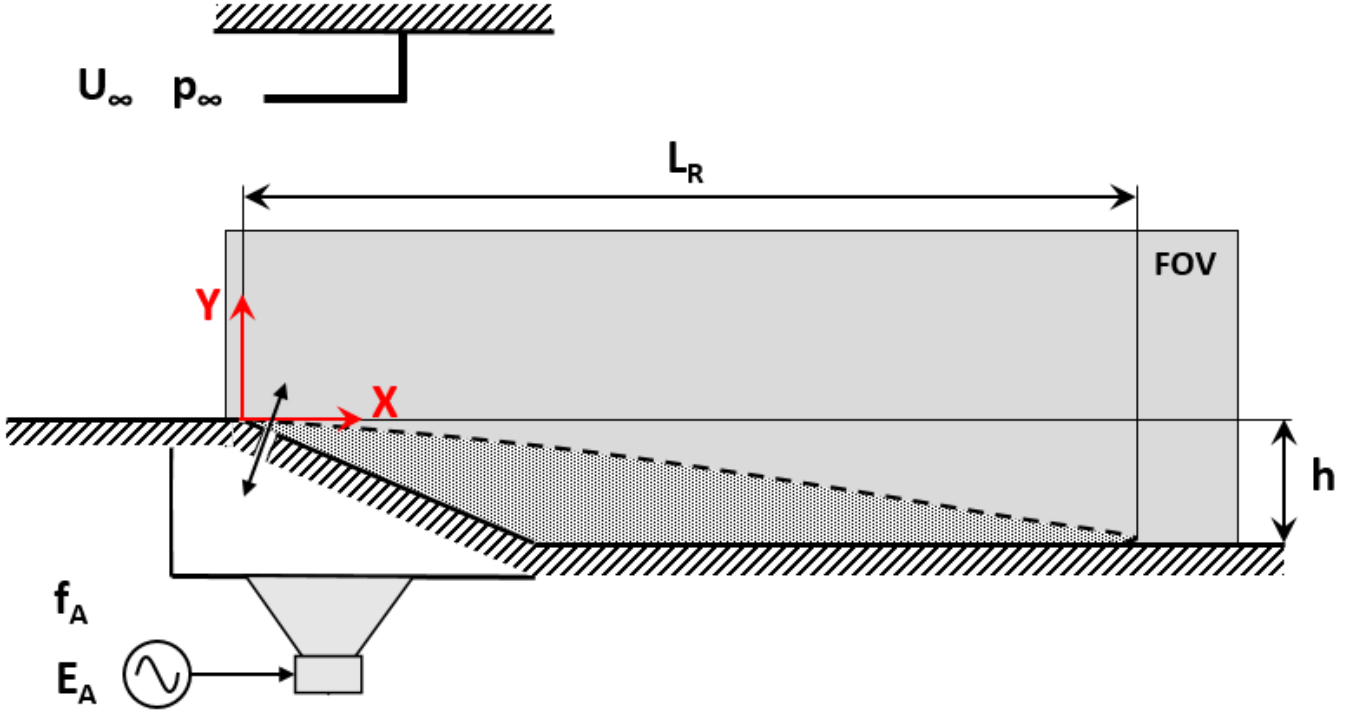}
  \caption{Sketch of the investigated model. The shaded area represents the recirculation region, while the PIV field of view (FOV) is visualised by a gray rectangle.}
  \label{fig:rampaGDR}
\end{figure}

We consider the massive turbulent separation that develops downstream of a salient edge, descending ramp of constant slope $\alpha = $ \SI{25}{\degree} and height $h =$ \SI{100}{\milli\metre}. A schematic view of the ramp as well as the reference system used throughout the experiment are given in Figure \ref{fig:rampaGDR}. Readers are referred to \citet{kourta15} and \citet{debien2016} for further details on the model. Measurements are taken in the S1, closed-loop wind tunnel at PRISME Laboratory, University of Orl\'{e}ans, France. The main test section is \SI{2}{\metre} by \SI{2}{\metre} wide and \SI{5}{\metre} long. The maximum reachable free stream velocity is \SI{60}{\metre\per\second}, with a turbulence intensity lower than \SI{0.3}{\percent}.  The model is placed at mid-height, and it spans the entire width $w$ of the test section. This provides an aspect ratio $w/h =$ \num{20}, which according to \citet{eatonJohnson1981} should be high enough to guarantee that the mean separated flow is bidimensional (see \cite{kourta15} for the baseline flow). 

\begin{figure}
  \centering
  \includegraphics[width=\linewidth]{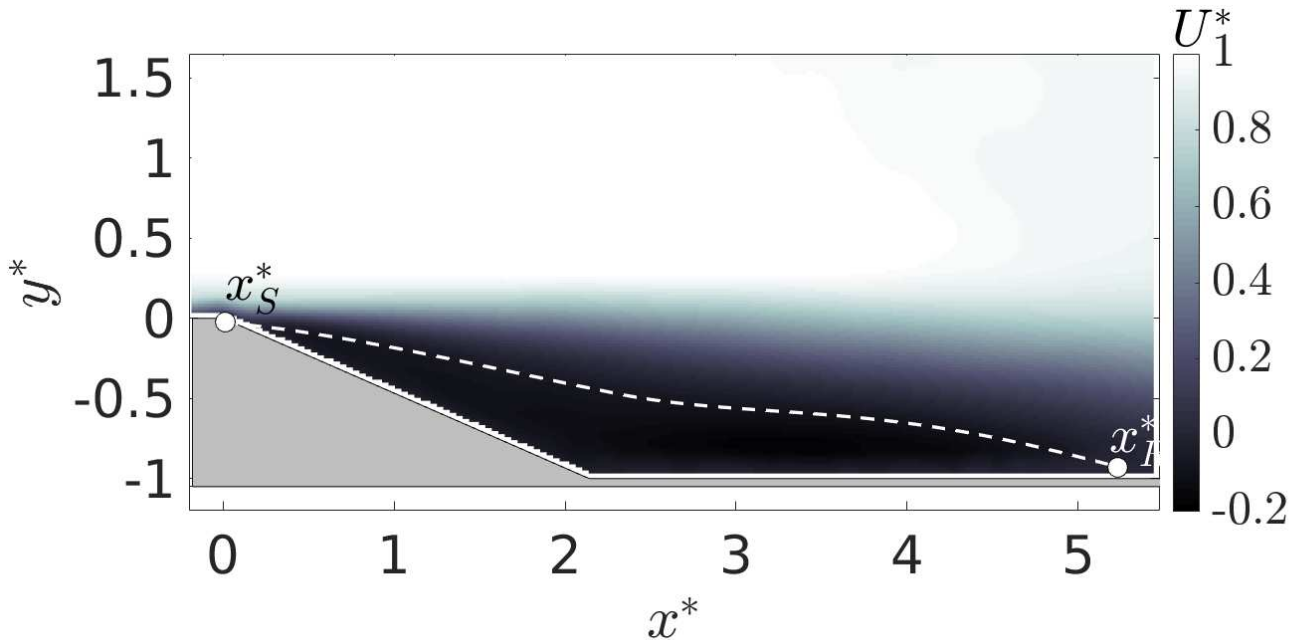}
  \caption{Streamwise velocity field of the baseline flow at $\Rey_\theta = $ \num{2550} ($U_\infty =$ \SI{20}{\meter\per\second}). The white dashed line (\protect\dashedrule) indicates the mean separation line.}
  \label{fig:Ufield_Unperturbed}
\end{figure}

Throughout the experiment, the reference velocity $U_\infty$, measured above the upper edge of the ramp, is fixed at \SI{20}{\meter\per\second}.
Following the prescriptions of \citet{adamsJohnstonPart1}, we use the parameter $\Rey_\theta$ to assess the intensity of turbulence in the incoming boundary layer. The chosen value of $U_\infty$ gives $\Rey_\theta \approx$ \num{2550}. To allow comparison with \citet{stella2017}, $\Rey_\theta$ is measured at $x/h = -9; y/h = 0$, by integrating the mean streamwise velocity profile \cite{BLTheory}. At the upper edge of the ramp, it is $\delta_e/h \approx 0.3$, where $\delta_e$ is the local thickness of the incoming boundary layer. Figure \ref{fig:Ufield_Unperturbed} presents the streamwise velocity field for the baseline separating/reattaching flow. The mean separation point $x_S^*$ is fixed at the upper edge of the ramp and the mean reattachment point is placed at $x_R^* \approx$ \num{5.23}, which is consistent with observations reported by \citet{kourta15}.

\subsection{Measuring devices}\label{sec:mesDev}
Since the mean flow is bidimensional, the characteristic parameters of the vortex model, which is the core of this study, can be recovered from 2D Particles Image Velocimetry (2D-2C PIV). Particle images are recorded at midspan, by two LaVision Imager LX 11M cameras (\SI[parse-numbers = false]{4032 \times 2688}{\square\pixel}). Taking into account the overlap between the fields of view (FOV) of each camera, the total FOV covers an area of \num{6.3}$h$ x \num{3.2}$h$ (see Figure \ref{fig:rampaGDR}). Since the spanwise vortex is a feature of the mean field, in this work we focus on first order statistics of the flow. Then, \num{500} independent image pairs for each tested control configuration are sufficient to attain adequate statistical convergence. Laser light is provided by a \SI{200}{\milli\joule}/\SI{15}{\hertz}/\SI{532}{\nano\metre} Quantel Evergreen 200-10 Nd:YAG laser, illuminating olive oil particles used to seed the flow. Images are correlated with the FFT, multi-pass algorithm of the LaVision Davis 8.3 software suit. The interrogation window is initially set to \SI[parse-numbers = false]{64 \times 64}{\square\pixel} and then reduced to \SI[parse-numbers = false]{32 \times 32}{\square\pixel}, with a \SI{50}{\percent} overlap. These settings yield a final vector spacing of \SI{1.89}{\milli\metre}. 

Pointwise velocity signals are acquired by single-component, hot-wire probes. Dantec 55P15 and 55P11 probes are respectively used for collecting velocity profiles in the incoming boundary layer and for characterising the response of the fluidic actuators. Signals are sampled at $f_s =$ \SI{60}{\kilo\hertz} for approximately \SI{10}{\second}. They are subsequently low-pass filtered with a cutoff frequency $f_c =$ \SI{30}{\kilo\hertz}.

Pressure at the lower edge of the ramp (see Figure \ref{fig:vortex}) is acquired by means of a Chell $\mu$DAQ-32C pressure scanner connected to a digital acquisition unit. The scanner has a full range (FR) of \SI{2.5}{\kilo\pascal} and an uncertainty of \SI{0.25}{\percent} FR. Pressure fluctuations are strongly damped by the pneumatic link between the pressure tap and the scanner \cite{kourta15}. Accordingly, only mean pressure variations are investigated, by averaging \num{3e4} samples acquired at $f_s =$ \SI{500}{\hertz}.

\section{Flow forcing and its effects on scaling}\label{sec:actuDev}
The ability of the vortex model to represent mean separated flows needs to be tested in many different flow configurations, for example by varying ramp geometry, or parameters of the flow such as $\Rey_\theta$. This task is often unpractical for the experimentalist: it requires additional measurements and time-consuming resets of the experimental set-up, while achieving flow modifications which, depending on the available facility and its constraints, are in many cases too small to validate the model on a sufficiently wide range of conditions. 
For example, in \citet{stella2017} a variation of almost half a decade of $\Rey_\theta$ leads to a reduction of $L_R^*$ of only \SI{20}{\percent}, and to modifications of mean flow topology which are mild, at least in the perspective of assessing the vortex model.

\begin{figure}
  \centering
  \includegraphics[width=\linewidth]{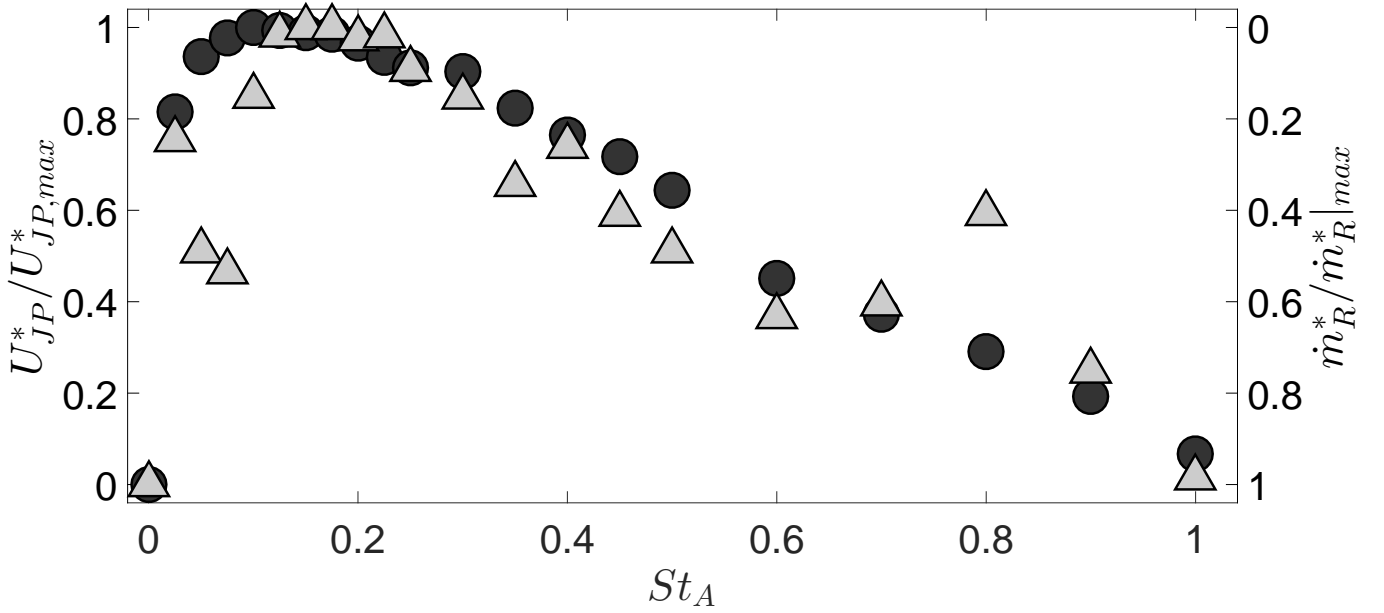}
\caption{Frequency response of the synthetic jet peak velocity $U_{JP}^*$ (\protect\blackcircle) and of the backflow $\dot{m}_R^*$ (\protect\graytriangleup).}
\label{fig:Freq_resp}
\end{figure}

A more efficient, alternative approach consists in artificially tuning the properties of the flow with an external forcing. Periodic actuators appear to be particularly well suited for this purpose, since a wide range of different flow configurations can be obtained simply by varying the actuation frequency (see references in Sec. \ref{sec:introVortex}). 
In this study, the flow is forced with a spanwise rack of 3 synthetic jets, designed and integrated in accordance with previous experimental and numerical studies \cite{kourta2013,guilmineau2014}. 
Each jet is ejected through a continuous slot of width $b = $ \SI{1}{\milli\metre} and length $l = $ \SI{660}{\milli\metre}, placed \SI{2}{\milli\meter} downstream of the upper edge of the ramp (see Figure \ref{fig:rampaGDR}). The jet is generated by a loudspeaker (Precision Devices PD.1550) and a cavity, fitted underneath the ramp. The axis of the jet is normal to the surface of the ramp. The loudspeakers are driven by single tone excitation signals in the form:
\begin{equation}
	E(t) = E_A \sin\left(2\pi f_A t\right),
	\label{eq:voltage}
\end{equation}
with $E_A$ being the peak-to-peak excitation voltage and $f_A$ the actuation frequency. In this study, $E_A$ is set to \SI{2}{\volt}, while $f_A$ is tuned within the range $\left[0,200\right]$ \si{\hertz}. Overall, \num{21} values of $f_A$ are tested.
For convenience, in the remainder of the paper $f_A$ will be expressed as a Strouhal number $St_A = f_A h/U_\infty$. 
The peak jet velocity $U_{JP}^*$, which depends non-linearly on the choice of $E_A$ and $St_A$, is characterised with a hot-wire probe. Measurement settings described at Sec. \ref{sec:mesDev} guarantee that the frequency response of the probe is well above the largest actuation frequency tested in this study. 
Figure \ref{fig:Freq_resp} shows that the actuator acts like a band-pass filter: its frequency response is almost flat at $U_{JP,max}^* \approx 2$ for $St_A \in \left(0.1;0.25\right)$ and drops significantly out of this range.

\begin{figure}
  \centering
  \includegraphics[width=\linewidth]{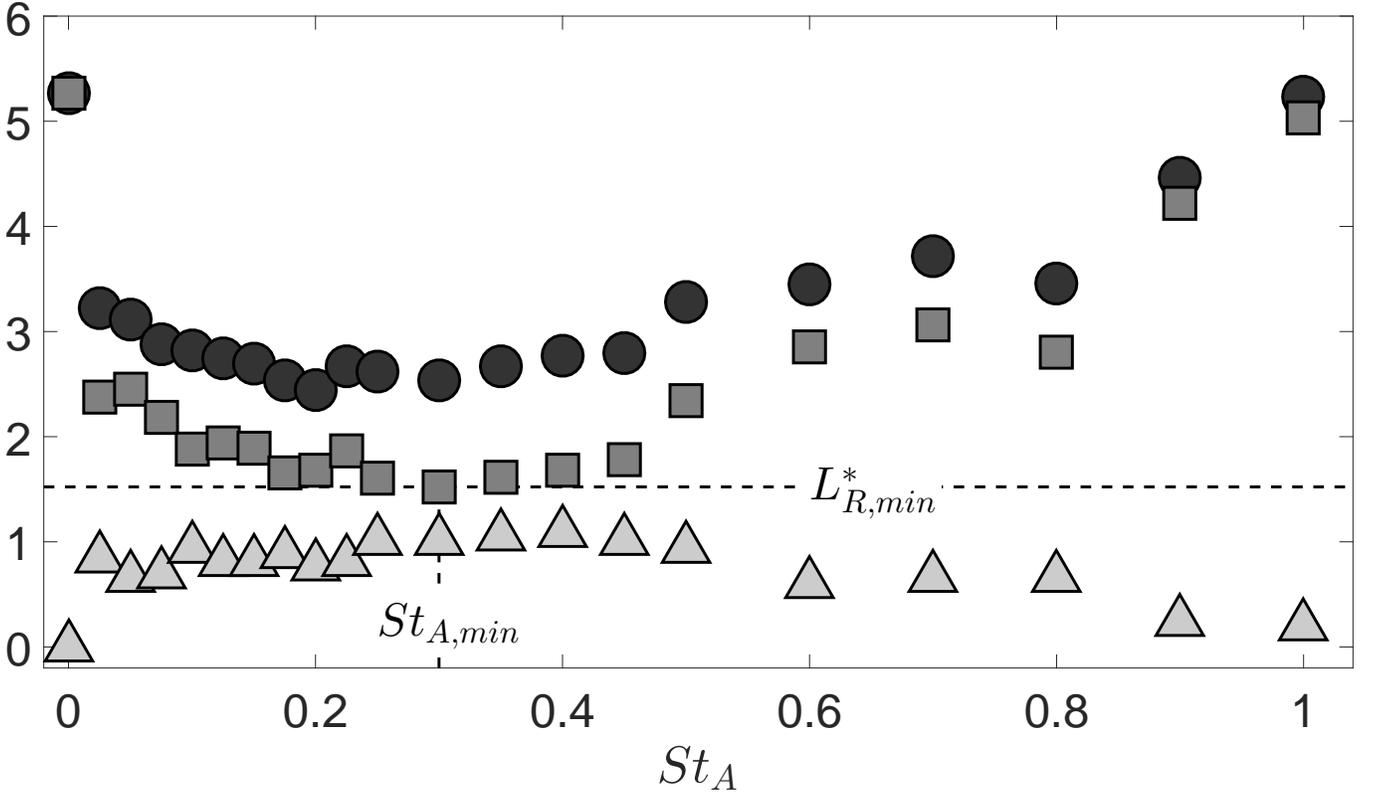}
\caption{Evolution with respect to $St_A$ of $L_R^*$ (\protect\graysquare), $x_S^*$ (\protect\graytriangleup) and $x_R^*$ (\protect\blackcircle).}
\label{fig:LR_StA}
\end{figure}

\begin{figure}
  \centering
  \includegraphics[width=\linewidth]{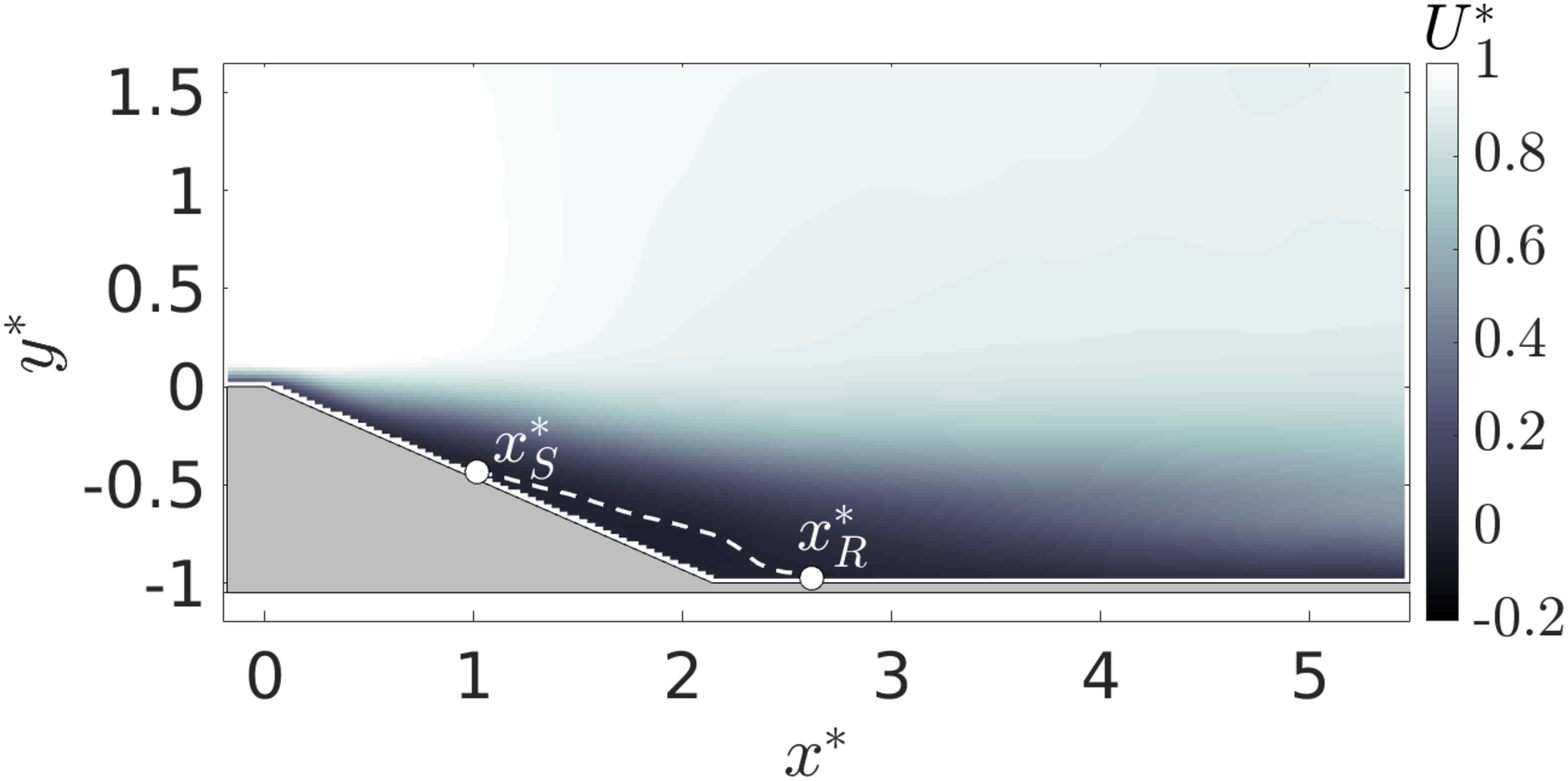}
  \caption{Streamwise velocity field of the controlled flow at $St_A =$ \num{0.25}. Symbols as in Figure \ref{fig:Ufield_Unperturbed}.}
  \label{fig:Ufield_controlled}
\end{figure}

Figure \ref{fig:LR_StA} presents the evolution of $L_R^*$ with respect to $St_A$. Results are in good agreement with previous observations \cite{shimizu1993,chun1996,berk2017}. Indeed, $L_R^*$ decreases strongly at low actuation frequencies, reaching a minimum $L_{R,min}^* \approx 1.5$ (i.e. as much as a \SI{75}{\percent} reduction of its baseline value) for $St_{A,min} \approx $ \numrange{0.25}{0.35}. 
$St_{A,min}$ corresponds to $St_L = St_A L_R^* \approx$ \numrange{0.45}{0.55}, which matches relatively well the natural shedding frequency of the separated shear layer, estimated at $St_L \approx$ \numrange{0.53}{0.66} by \citet{debien14}. For $St_A>St_{A,min}$, $L_R^*$ increases once again, recovering its baseline value for $St_A \approx 1$. 
Generally speaking, the variation of $L_R^*$ depends on the displacements of both the mean reattachment point $x_R^*$ and of the mean separation point $x_S^*$. Figure \ref{fig:LR_StA} shows that the trends of $x_S^*$ and $x_R^*$ with respect to $St_A$ are quite well correlated with each other and with the evolution of $L_R^*$. For example, a reduction of $L_R^*$ is due to both $x_R^*$ moving upstream and to $x_S^*$ being displaced downstream by the action of the synthetic jets. 
For comparison with the baseline flow (Figure \ref{fig:Ufield_Unperturbed}), Figure \ref{fig:Ufield_controlled} reports an example of the mean topology of a controlled flow.

\begin{figure}
  \centering
  \includegraphics[width=\linewidth]{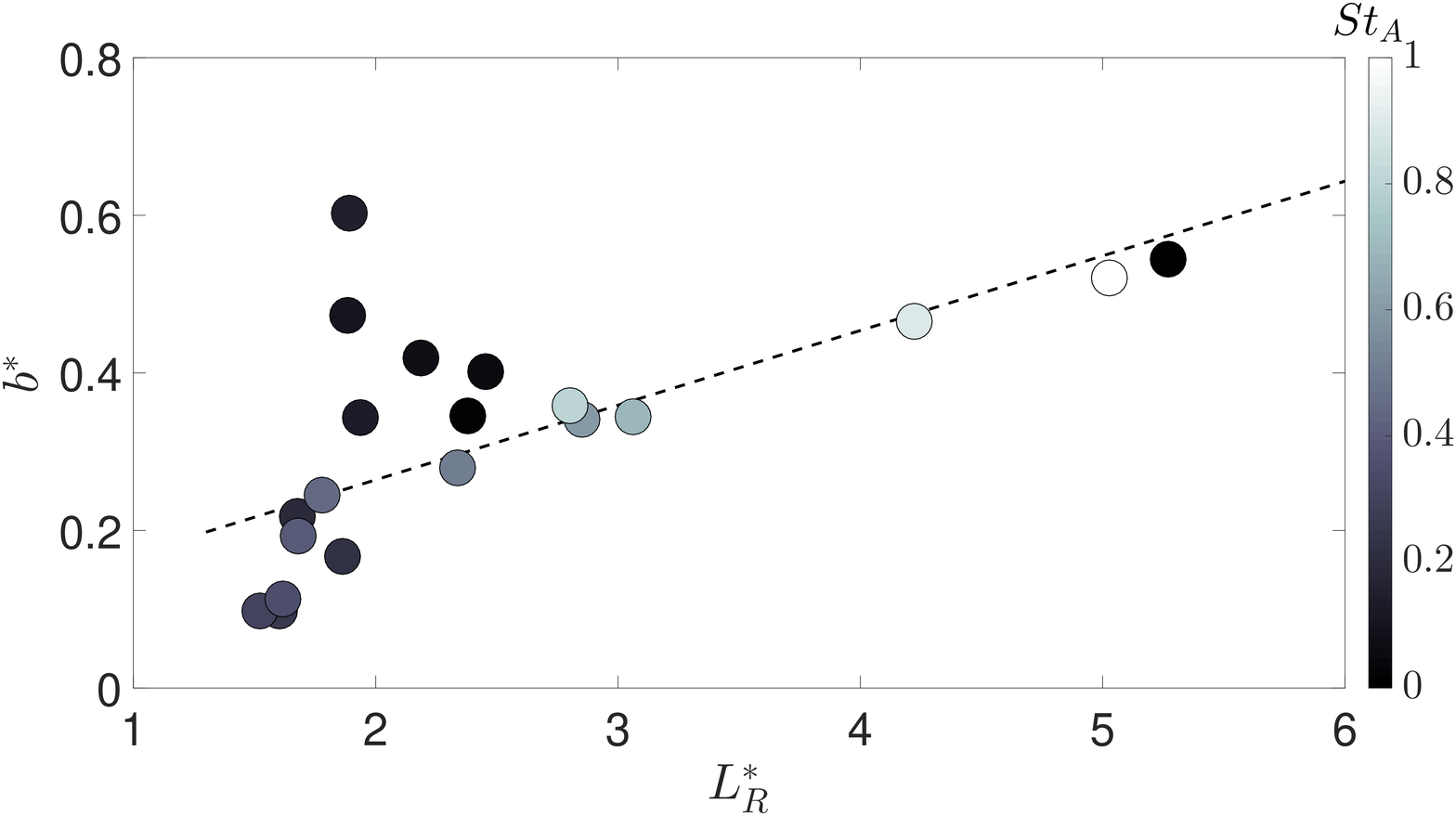}
\caption{Correlation between $L_R^*$ and $b^* \approx \left|-1 - y_V^*\right|$. The grayscale indicates the value of $St_A$ and the dashed line (\protect\dashedrule) represents a linear best-fit of available data.}
\label{fig:yV_LR}
\end{figure}

\begin{figure}
  \centering
  \includegraphics[width=\linewidth]{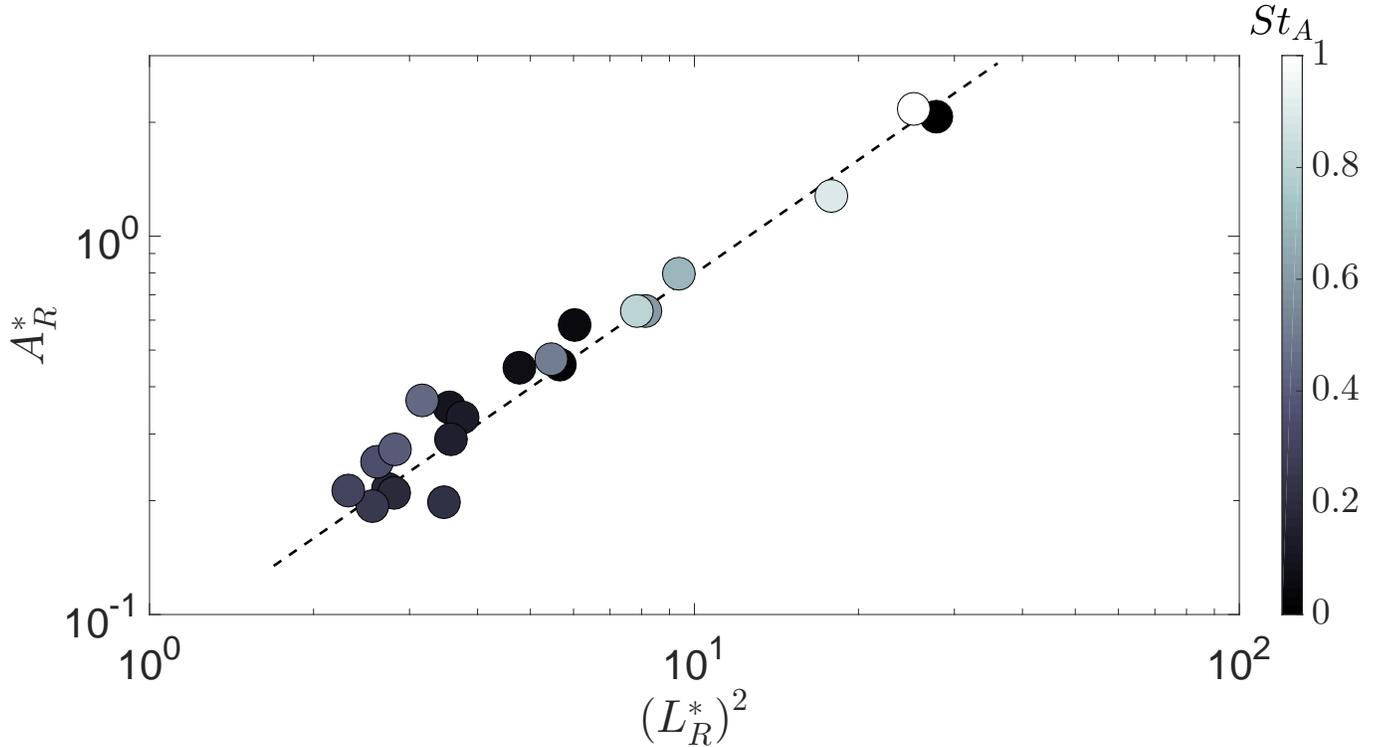}
\caption{Correlation between $(L_R^*)^2$ and $A_R^*$. The grayscale indicates the value of $St_A$ and the dashed line (\protect\dashedrule) represents a best-fit of Eq. \ref{eq:AR_scaling_2} on available data.}
\label{fig:AR_LR}
\end{figure}

The topological definition of the vortex model given in introduction links the size of the vortex to $L_R^*$, implicitely considering that this is the main characteristic scale of the mean flow. 
This is generally an acceptable assumption for a natural (i.e. non forced) flow. Anyway, \citet{berk2017} show that the synthetic jets introduce an additional macroscopic scale into the \textit{instantaneous} flow, which is the $St_A$-dependent characteristic length scale of the train of vortices that they generate.
Before going any further in the discussion of the vortex model, then, it seems important to verify whether this new scale modifies the scaling of the mean flow, or if $L_R^*$ remains the dominant length scale of controlled cases too.
One simple starting point to investigate this matter might be found by considering how the external forcing affects the surface (i.e. the volume per unit spanwise length) of the vortex. With reference to Figure \ref{fig:vortex}, let us approximate the shape of the vortex with an ellipse. Then, it will be:
\begin{equation}
A_V^* \approx \pi a^* b^*,
\label{eq:Av_2}
\end{equation}
in which $a^*$ and $b^*$ are the semi-major axis and the semi-minor axis of the ellipse, respectively. Figure \ref{fig:Ufield_Unperturbed} and Figure \ref{fig:Ufield_controlled} suggest that the major axis can be associated with the mean separation line, so that $a^* \sim L_R^*$.
This being so, it does not seem unreasonable to postulate that, in first approximation, the external forcing will at most affect the scaling of $b^*$.
Let us then study the evolution of $b^*$ in more details.
Since $\alpha$ is not too large, Figure \ref{fig:vortex} suggests to put $b^* \approx \left|-1 - y_V^*\right| = 1 + y_V^*$, where $y_V^*$ is the vertical position of the center of the vortex.
As a feature of the time-averaged flow, the vortex is stationary. Then, its center can be identified as a point of the mean separation line where $V=0$, placed in the neighbourhood of $(x^*-x_S^*)/L_R^* = 0.5$.
Figure \ref{fig:yV_LR} shows the trend of $b^*$ with respect to $L_R^*$. Although scatter is not negligible for $L_R^* <2.5$, most available datapoints contribute to sketch a clear linear correlation between $b^*$ and $L_R^*$, which encourage us to put:
\begin{equation}
b^* \approx \left(1 + y_V^*\right) \sim L_R^*.
\label{eq:yV_LR}
\end{equation} 
and hence:
\begin{equation}
A_V^* \sim \left(L_R^*\right)^2.
\label{eq:AR_scaling_2}
\end{equation}
Direct computation of $A_V^*$ is, in general, not straightforward. Anyway, the surface of the recirculation region $A_R^*$ seems to be an adequate estimator of $A_V^*$. Indeed, on the one hand the definition of the vortex suggests that $A_V^* \approx 2 A_R^*$ (see Sec. \ref{sec:introVortex}). On the other hand, $A_R^*$ is delimited by the mean separation line (i.e. $\approx 2a^*$) while $b^*$ approximately corresponds to its height. 
This being so, Figure \ref{fig:AR_LR} reports the evolution of $A_R^*$ with respect to $\left(L_R^*\right)^2$, showing that available data agree well with Eq. \ref{eq:AR_scaling_2}.
All in all, these results imply that $L_R^*$ remains the main characteristic length scale of the mean separating/reattaching flow, regardless to the working point of the actuators. The scales typical of the train of vortices generated by the synthetic jets \cite{berk2017} do not seem to affect the scaling of the mean flow.
Hence, the definition of the vortex model given at Sec. \ref{sec:introVortex} is relevant to the study of forced separating/reattaching flows.

\section{Mass entrainment}\label{sec:LR_backflow}
The objective of this section is to illustrate how the vortex model takes into account mass exchanges within the separated flow and, in particular, how its defining parameters $\Gamma_V^*$ and $L_R^*$ can be associated to the backflow $\dot{m}_R^*$.
\subsection{Computation of the backflow}\label{sec:backflowComp}
The first necessary step of our investigation is to evaluate the intensity of the backflow and its evolution with $St_A$. 
It is once again reminded that $\dot{m}_R^*$ is the mass flux that goes through the recirculation region \cite{stella2017}.
Interestingly, the recirculation region has only one permeable boundary, that is the mean separation line. 
Since the mean flow is bidimensional, continuity implies that the total mass flux through the mean separation line should be zero, that is:
\begin{equation}
-\frac{1}{U_\infty h}\int_{S_R}V cos\left(\phi\right) ds = 0,
\label{eq:massBalance}
\end{equation}
where $s$ is a curvilinear abscissa defined along the mean separation line $S_R$ and $\phi$ is the angle between the local normal (pointing toward the free flow) and the $Y$ axis.
\begin{figure}
  \centering
  \includegraphics[width=\linewidth]{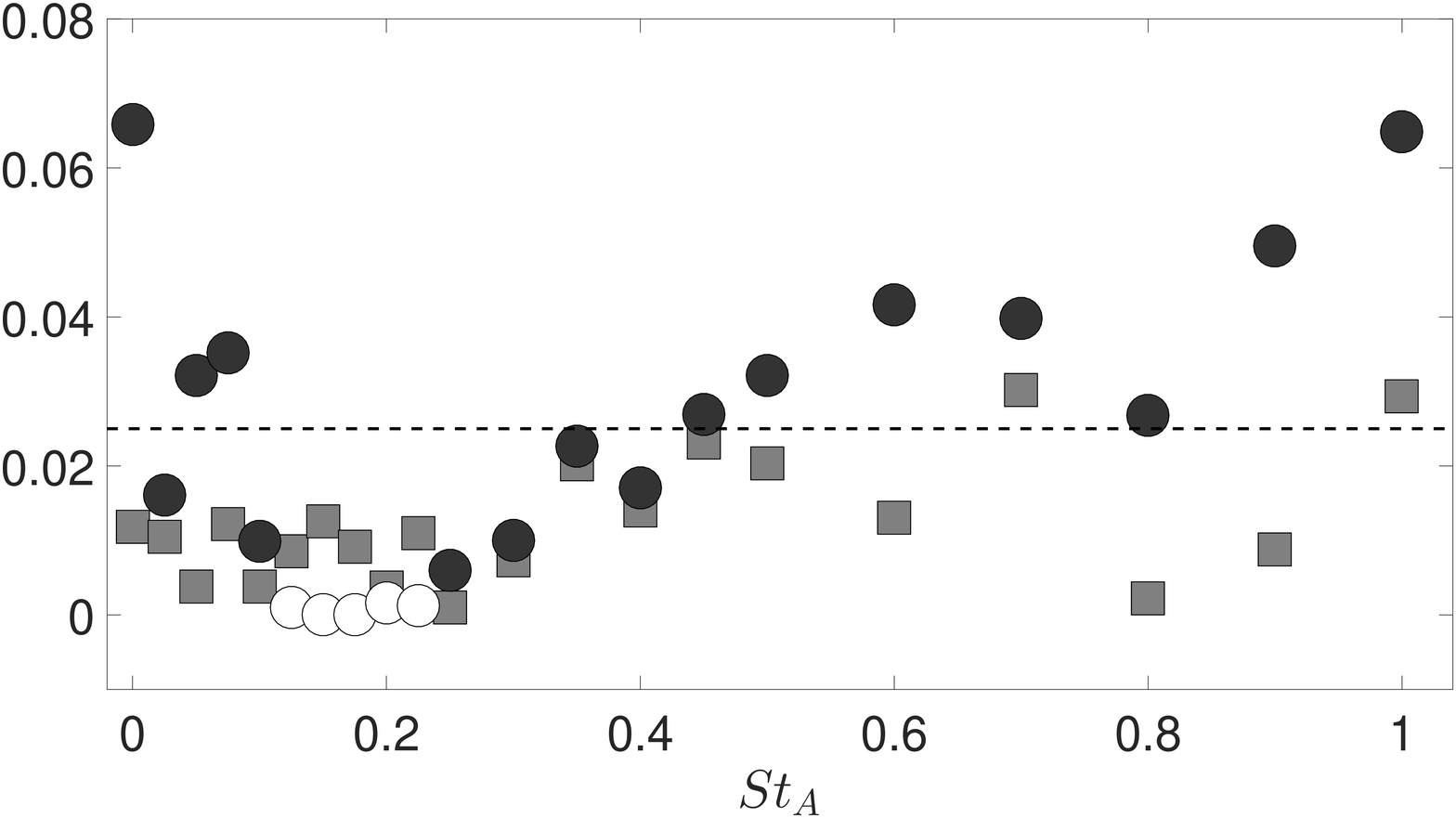}
\caption{Mass balance along the mean recirculation region. Symbols: \protect\graysquare $\left|\epsilon_R^*\right|$; \protect\dashedrule $\left|\epsilon_R^*\right| = 0.025$; \protect\whitecircle $\dot{m}_R^*$. Empty symbols indicate points for which $\left|\epsilon_R^*\right| < \dot{m}_R^*$.}
\label{fig:mass_error}
\end{figure}
As shown in Figure \ref{fig:mass_error}, the recirculation region verifies Eq. \ref{eq:massBalance} approximately within an error $\epsilon_R^* \approx \pm$ \num{2.5e-2} for all tested values of $St_A$, which seems acceptable, at least in comparison to errors reported by \citet{stella2017}. 
A large fraction of $\epsilon_R^*$ seems to be due to laser reflections at the wall, that produce spurious vectors in regions of the velocity field that surround the extrema of the mean separation line. 

It is clear from Figure \ref{fig:vortex} that the amount of mass that rotates with the vortex should alone assure continuity. Eq. \ref{eq:massBalance} can then be recasted as:
\begin{equation}
\dot{m}_R^{* \, IN} + \dot{m}_R^{* \, OUT} = 0,
\label{eq:massBalance_2}
\end{equation}
where the superscripts $^{IN}$ and $^{OUT}$ indicate mass injection and extraction, respectively. According to Eq. \ref{eq:massBalance_2}, then, the estimation of the backflow $\dot{m}_R^*$ requires to condition the velocity integral of Eq. \ref{eq:massBalance} to the sign of $V$, which gives:
\begin{equation}
\dot{m}_R^* = \dot{m}_R^{* \, IN} = - \frac{\int_{S_R}V^{-}cos\left(\phi\right) ds}{U_\infty h} = - \dot{m}_R^{* \, OUT} = \frac{\int_{S_R}V^{+}cos\left(\phi\right) ds}{U_\infty h}.
\label{eq:backflow}
\end{equation}
In this expression, $V^{+}$ represents positive velocity values and $V^{-}$ negative ones, respectively.
In itself, continuity does not put any constraint on the form of the integrands of Eq. \ref{eq:backflow}. Anyway, results reported by \citet{stella2017} for the unperturbed flow show that $V^{+}$ are rather concentrated over $x/h < L_R^*/2$, while $V^{-}$ are found on $x/h > L_R^*/2$. This is also well verified in the present experiment, at least if $L_R^*$ is not too small, as suggested by shape of the streamlines shown in Figure \ref{fig:vortex}.
In addition, phenomenological arguments supporting such an approximately odd distribution of $V$ along the mean separation line can be found in the work of \citet{chapman1958}: these researchers highlighted that the amount of mass \textit{scavenged} into the recirculation region at reattachment should balance the flux of mass re-entrained into the shear layer at separation. 
The evolution of $\dot{m}_R^*$ with $St_A$, computed using Eq. \ref{eq:backflow}, is presented in Figure \ref{fig:Freq_resp} and Figure \ref{fig:mass_error}. 
Interestingly, Figure \ref{fig:Freq_resp} shows that $\dot{m}_R^*$ is approximately inversely correlated to the frequency response of the jet, even if the minimum value of $\dot{m}_R^*$ seems to be attained for higher values of $St_A$ than those for which $U_{JP}^*$ reaches its maximum. 
In any case, it appears that, in first approximation, the higher is $U_{JP}^*$, the more effective is the jet at hindering the backflow.

It is important to remark that even if the value of $\epsilon_R^*$ is quite homogeneous across the spanned $St_A$ domain, its relative impact on our investigation increases as $\dot{m}_R^* \rightarrow 0$. In particular, it is $\epsilon_R^* > \dot{m}_R^*$ on $St_A \in (0.12,0.22)$. For sake of safety, then, datapoints within this frequency subrange (highligted by empty symbols in Figure \ref{fig:mass_error}) will be discarded in the remainder of the paper.

\subsection{Backflow and vortex circulation}\label{sec:backflowCirc}
\begin{figure}
  \centering
  \includegraphics[width=\linewidth]{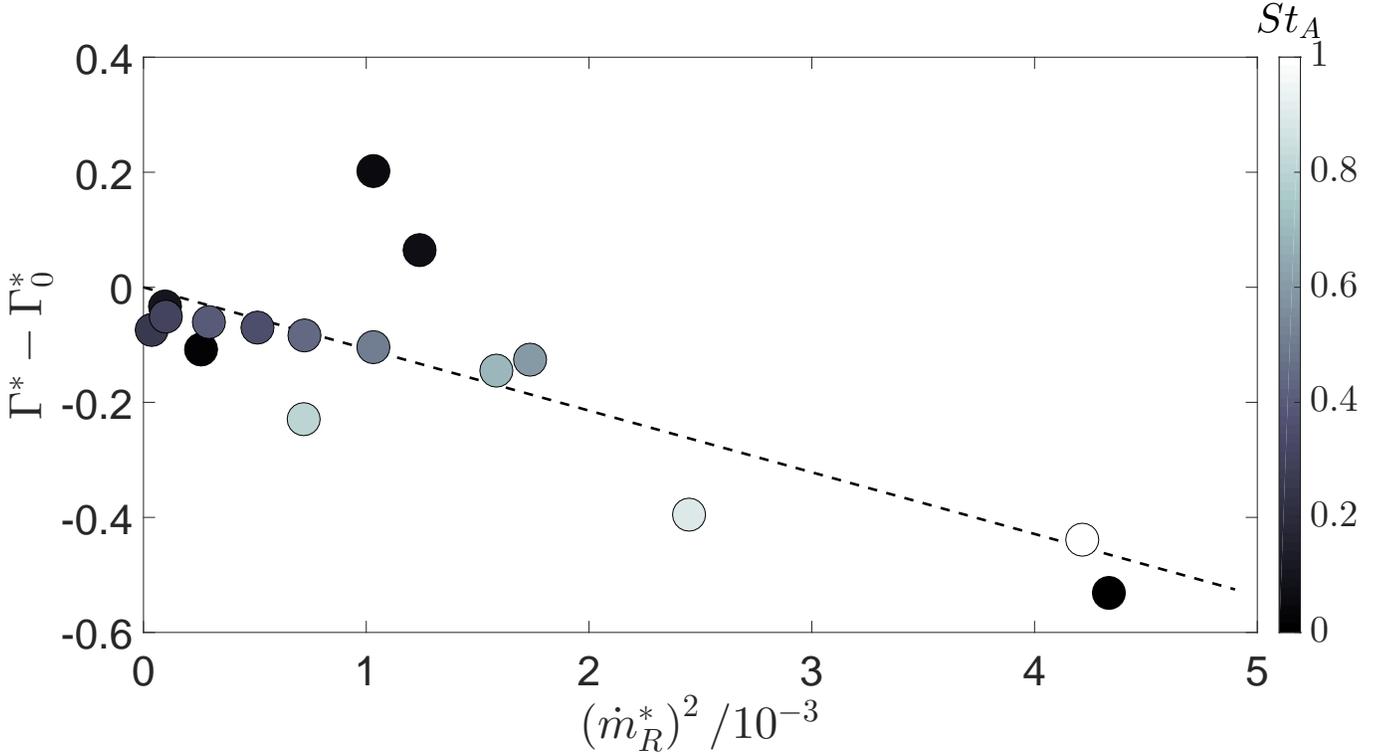}
\caption{Evolution of $\Gamma^* - \Gamma_0^*$ with respect to $\left(\dot{m}_R^*\right)^2$. The grayscale indicates the value of $St_A$ and the dashed line (\protect\dashedrule) represents a best-fit of Eq. \ref{eq:gamma_mR} on available data.}
\label{fig:GammaV_mR}
\end{figure}

Now that values of $\dot{m}_R^*$ are available, we can start the discussion of the vortex model by investigating whether $\dot{m}_R^*$ can be related to $\Gamma_V^*$.
Let us begin by making the assumption that the total circulation of the flow can be modelled as:
\begin{equation}
\Gamma^* = \Gamma_0^* + \Gamma_V^*,
\label{eq:gamma_Tot}
\end{equation}
where $\Gamma_0^*$ is the circulation of a hypothetical flow with no separation. 
In principle, $\Gamma_0^*$ should only depend on geometry and free flow velocity. In the reference system of Figure \ref{fig:rampaGDR}, it is both $\Gamma_0^* < 0$ and $\Gamma_V^*<0$.
Eq. \ref{eq:gamma_Tot} implies that all variations of $\Gamma^*$ measured in the experiment should be attribuable to variations of the properties of the vortex. In this respect, it does not seem unreasonable to put: 
\begin{equation}
\Gamma_V^*(St_A) \sim \Gamma_{jet}^*(St_A),
\label{eq:gammaV_gammaJ}
\end{equation}
as the size of the vortex is driven by the action of the synthetic jets (see Figure \ref{fig:LR_StA}). Mind that the correlation postulated in Eq. \ref{eq:gammaV_gammaJ} appears to be a negative one. Indeed, since the velocity scale of the free flow remains $U_\infty$ (see also Sec. \ref{sec:wallNormPrexGrad} for what concerns the recirculation region), $\Gamma_V^*$ should decrease as $L_R^*$ shrinks and hence, according to the analysis proposed by \citet{berk2017}, as $\Gamma_{jet}^*(St_A)$ increases. The same work by \citet{berk2017} suggests that the mean amount of circulation injected in the flow by the controller is given by:
\begin{equation}
\Gamma_{jet}^* \sim \frac{1}{T U_\infty^2}\int_0^T \frac{1}{2}U_J^2(t,St_A)\mathrm{d}\tau \sim (U_{JP}^*)^2 \sim \left(\dot{m}_R^*\right)^2,
\label{eq:gammaJet}
\end{equation}
in which $\tau$ indicates time, $T$ is the actuation period and the dependency on $St_A$ was omitted for simplicity. According to Figure \ref{fig:Freq_resp}, the proportionality with respect to $\dot{m}_R^*$ should hold at least on $St_A > 0.1$. By plugging Eq. \ref{eq:gammaJet} into Eq. \ref{eq:gamma_Tot} one obtains:
\begin{equation}
\Gamma^* - \Gamma_0^* = \Gamma_V^* \sim \left(\dot{m}_R^*\right)^2,
\label{eq:gamma_mR}
\end{equation}
In a bidimensional flow, $\Gamma^*$ is classically computed as:
\begin{equation}
\Gamma^* = \frac{1}{U_\infty h}\oint_{S_V} \boldsymbol{U}\cdot\boldsymbol{t} \, ds,
\label{eq:gammaDef}
\end{equation}
where $\boldsymbol{t}$ is the local tangent to the closed contour $S_V$, which is presented in Figure \ref{fig:vortex}. As for what concerns $\Gamma_0^*$, it can be estimated by fitting Eq. \ref{eq:gamma_mR} onto available values of $\Gamma^*$ and $\dot{m}_R^*$, which gives $\Gamma_0^* \approx -5.21$.
Figure \ref{fig:GammaV_mR} shows that, in spite of some scatter, computed values of $\Gamma_V^*$ verify Eq. \ref{eq:gamma_mR} relatively well. 
It seems then safe to consider that $\dot{m}_R^*$ is related to $\Gamma_V^*$, i.e. that the backflow can be assimilated to the amount of mass put in rotation by the vortex.
Since the vortex dominates the separated flow, the link with $\Gamma_V^*$ gives strong support to the idea that $\dot{m}_R^*$ is a key quantity in the functionning of separating/reattaching flows.

\subsection{Backflow and vortex size}\label{sec:LR_mR}
\begin{figure}
  \centering
  \includegraphics[width=\linewidth]{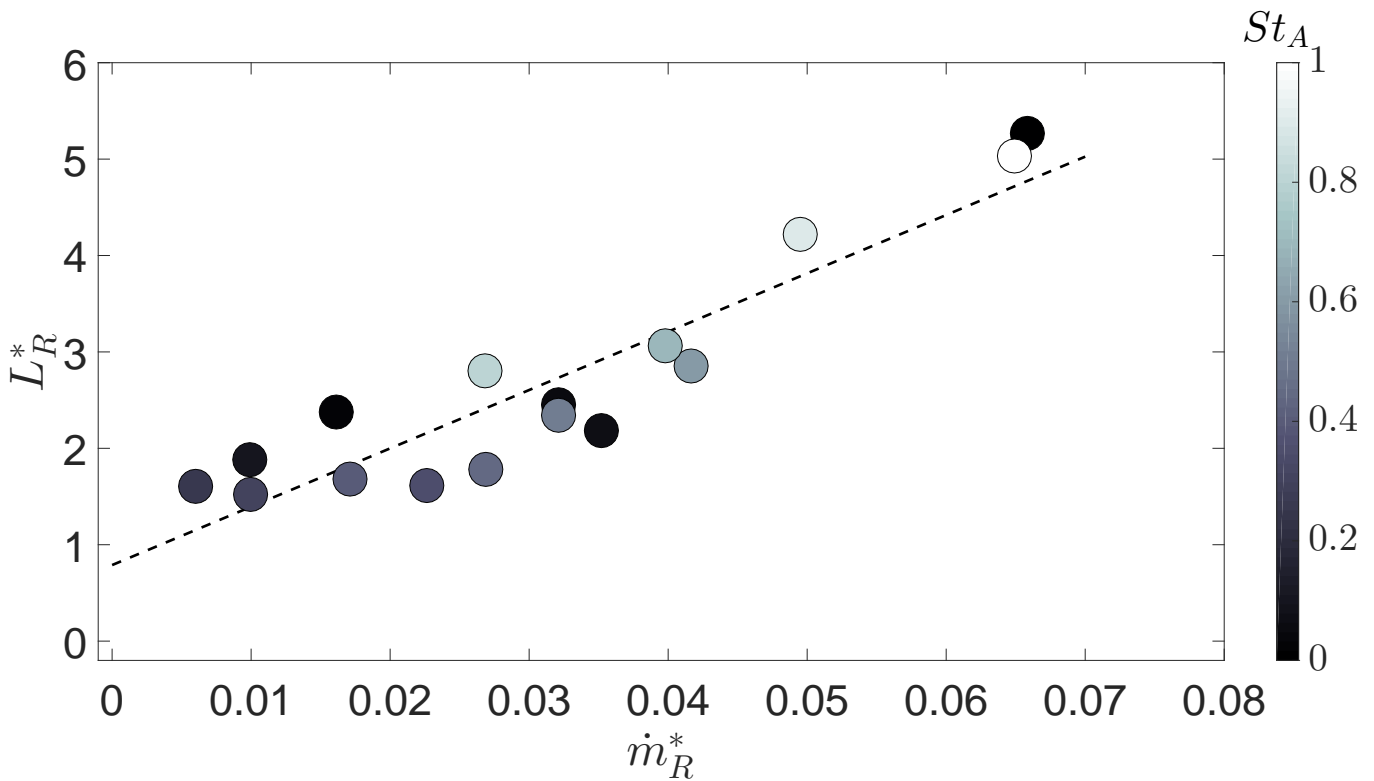}
\caption{Evolution of $L_R^*$ with respect to $\dot{m}_R^*$. The grayscale indicates the value of $St_A$ and the dashed line (\protect\dashedrule) represents a linear fit as in Eq. \ref{eq:LR_mR}.}
\label{fig:LR_mR}
\end{figure}

The second vortex parameter that has to be compared to $\dot{m}_R^*$ is its characteristic scale $L_R^*$.
Not surprisingly, Figure \ref{fig:LR_mR} shows that $\dot{m}_R^*$ and $L_R^*$ are directly correlated, as could be expected by the fact that $L_R^*$ scales the exchange surface of the mean recirculation region (see Eq. \ref{eq:backflow} and discussion). The trend of $L_R^*$ with respect to $\dot{m}_R^*$ appears to be well approximated by a linear model in the form:
\begin{equation}
L_R^* = k_{m,L}\dot{m}_R^* + L_{R,0}^*,
\label{eq:LR_mR}
\end{equation}
where $k_{m,L}$ and $L_{R,0}^*$ were estimated at approximately \num{60} and \num{0.8}, respectively, by fitting Eq. \ref{eq:LR_mR} on available data.
The non-null value of $L_{R,0}^*$ is somehow surprising, because it seems to imply that the recirculation region does not disappear for $\dot{m}_R^* \rightarrow 0$. 
In this paper we do not tackle this problem directly, but some arguments supporting $L_{R,0}^*>0$ are discussed in Appendix.

Regardless to the value of $L_{R,0}^*$, anyway, Eq. \ref{eq:LR_mR} is in fundamental agreement with findings reported by \citet{berk2017}. Indeed, comparison of Figure \ref{fig:LR_mR} with Figure \ref{fig:LR_StA} proves that $\dot{m}_R^*$ is a more appropriate predictor of the evolution of controlled separating/reattaching flows than $St_A$ \cite{berk2017}, providing a sensibly simpler, monotonical description of the behaviour of the characteristic length scale $L_R^*$.
It is stressed that the linear trend presented in Figure \ref{fig:LR_mR} was obtained without making any hypothesis on the characteristics of the actuator. In fact, observations reported by \citet{stella2017} on the scaling of $L_R^*$ in a unperturbed flow suggest that a linear relationship between $\dot{m}_R^*$ and $L_R^*$, similar to Eq. \ref{eq:LR_mR}, might be a general property of separating/reattaching flows assimilable to the one under study.

\section{An estimator for $\boldsymbol{\dot{m}_R^*}$}\label{sec:backflowObs}
By exploiting the vortex model, previous sections contributed at affirming the key role of mass entrainment in separating/reattaching flows. In this regard, two findings seem particularly promising in the perspective of separation control. In the first place, it appears that $L_R^*$ (i.e. the \textit{state} of the flow, at least for what concerns topological analysis) could be reconstructed simply from the backflow $\dot{m}_R^*$. 
In the second place, the inverse correlation between $\dot{m}_R^*$ and the characteristic actuation velocity $U_{JP}^*$ might hold promise of entrainment-based, closed-loop control systems, in which $\dot{m}_R^*$ feedback might be used to regulate the intensity of the command given to attain a target flow state.
In the light of these elements, in principle $\dot{m}_R^*$ stands out as a powerful control variable, that might provide both information to reconstruct the state of the flow and input for actuation.
Unfortunately, in practice measuring $\dot{m}_R^*$ is impossible in most real-life applications, because the large velocity fields that are necessary to its computation are usually not available (see Sec. \ref{sec:backflowComp}). 
In this respect, we argue that the vortex model might provide a model-based definition of simply deployable observers for $\dot{m}_R^*$, and hence $L_R^*$.

\subsection{Relating $\boldsymbol{\dot{m}_R^*}$ to the pressure field}\label{sec:backFlow2Prex}
Practical problems in sensing industrial flows make it suitable to base the estimation of $\dot{m}_R^*$ on simply accessible information. The first candidate quantity that comes to mind is wall-pressure, which in most applications can be directly measured with relatively unexpensive, available on-the-shelf, flush-mounted pressure taps.
The mean wall-pressure distribution typical of separating/reattaching flows assimilable to the one under study is well characterised for the baseline flow (for example, see \cite{roshko1965} and related, subsequent literature) and already documented in the case of flows controlled with synthetic jets \cite{guilmineau2014}. 
Anyway, a quantitative link between mean wall-pressure (and the pressure field more in general) and $\dot{m}_R^*$ is not self evident.
It is then convenient to begin our discussion by explicitely investigating if such link exists. 
In this regard, the vortex model can help our reasoning, as follows.
Since the vortex dominates the mean flow, it does not seem unreasonable to relate the vertical pressure force acting on the mean separation line to the circulation $\Gamma_V$. By invoking the Joukowski theorem, this relation can be expressed as:
\begin{equation}
\rho U_\infty\Gamma_V \approx \int_{S_R}\left(P(s)-P_\infty\right) ds \approx \left(P_V-P_\infty\right)S_R \sim \left(P_V-P_\infty\right)L_R,
\label{eq:Jukou1}
\end{equation}
in which $P(s)$ is the pressure distribution along the mean separation line, $P_V$ is its average value and $P_\infty$ is the mean static pressure in the free stream above the descending ramp. It suits our purposes to consider that $P_V \approx P(x_V,y_V)$, where $x_V$ is the streamwise position of the vortex center (see Figure \ref{fig:vortex}). At least for what concerns the baseline flow, this hypothesis is supported by the odd form of the pressure gradient reported by \citet{stella2017}. By normalising all quantities in Eq. \ref{eq:Jukou1}, one naturally obtains:
\begin{equation}
\Gamma_V^* \sim C_{p,V}L_R^*.
\label{eq:Jukou3}
\end{equation}
In this expression, $C_{p,V}$ can be interpreted as the characteristic pressure coefficient of the center of the vortex, computed as:
\begin{equation}
C_{p,V} = \frac{P_V - P_\infty}{1/2 \rho U_\infty^2}.
\label{eq:Cp}
\end{equation}
Starting from the vortex model, in previous sections we have already proposed dependencies on $\dot{m}_R^*$ for both $\Gamma_V^*$ (Eq. \ref{eq:gamma_mR}) and $L_R^*$ (Eq. \ref{eq:LR_mR}). By plugging these expressions into Eq. \ref{eq:Jukou3}, one simply obtains:
\begin{equation}
C_{p,V} \sim \frac{\Gamma_V^*}{L_R^*} \sim \dot{m}_R^*,
\label{eq:vortexPrex}
\end{equation}
which should hold at least if $L_R^*$ is not too small with respect to $L_{R,0}^*$. 
Eq. \ref{eq:vortexPrex} provides a first connection between the pressure field and the backflow, even if $C_{p,V}$ is generally not accessible without deeply perturbing the separated flow. We then need to go one step further in our reasoning, and relate $C_{p,V}$ to a wall-pressure value.
 
\subsection{The wall-normal pressure gradient within the recirculation region}\label{sec:wallNormPrexGrad}
\begin{figure}
  \centering
  \includegraphics[width=\linewidth]{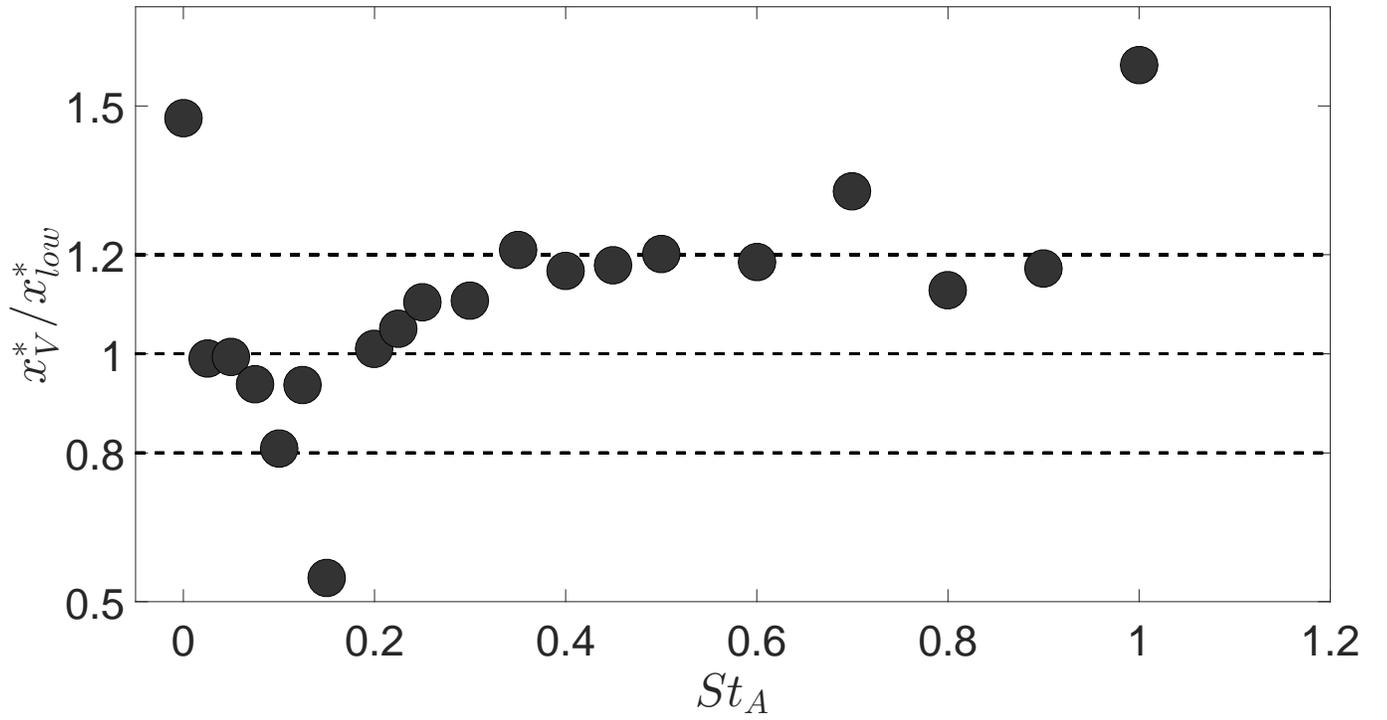}
\caption{Evolution of the streamwise position of the vortex $x_V^*$ with respect to $St_A$. All available datapoints are included, regardless to the value of $\epsilon_R^*$ (see Sec. \ref{sec:backflowComp}).}
\label{fig:xV_Sth}
\end{figure}

\begin{figure}
  \centering
  \includegraphics[width=\linewidth]{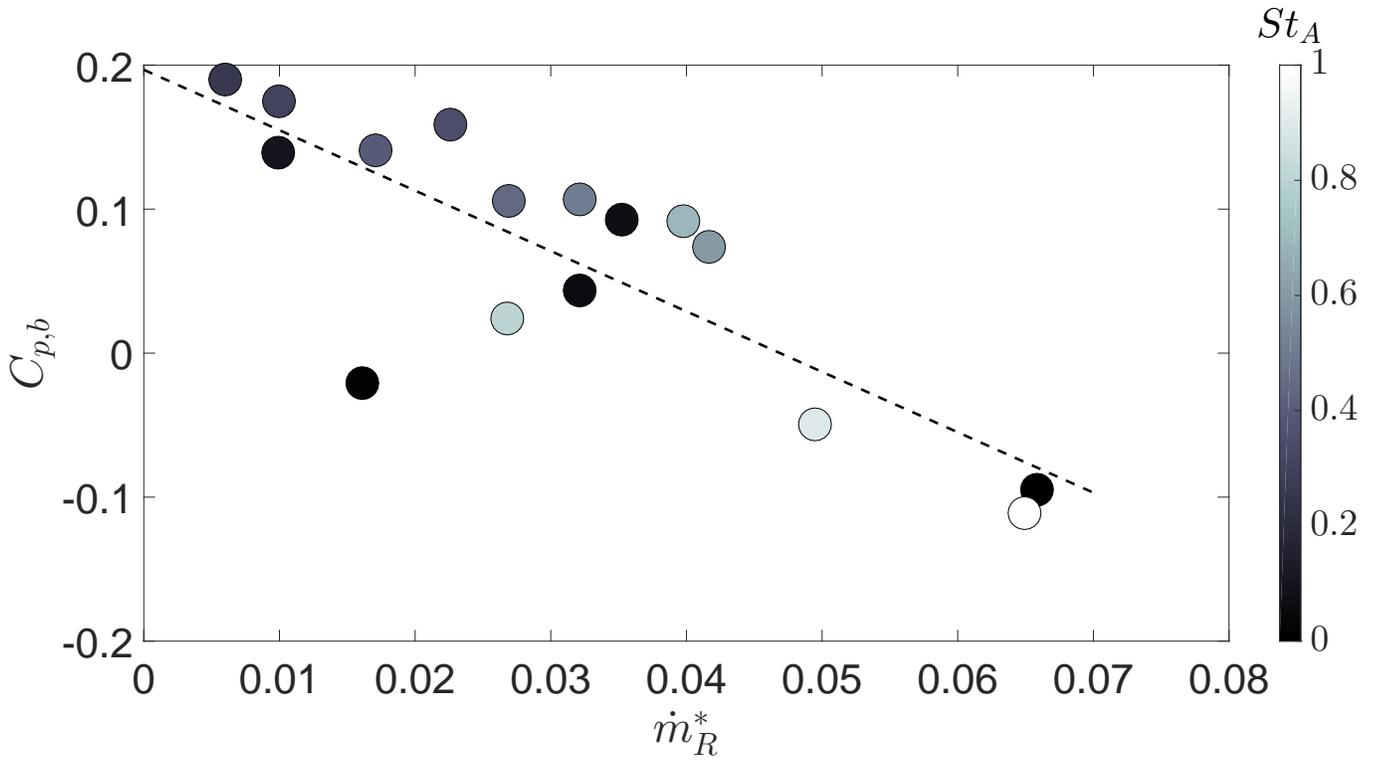}
\caption{Evolution of $C_{p,b}$ with respect to $\dot{m}_R^*$. The grayscale indicates the value of $St_A$ and the dashed line (\protect\dashedrule) represents a linear fit.}
\label{fig:Cp_mR}
\end{figure}

In order to introduce mean wall-pressures into Eq. \ref{eq:vortexPrex}, it is useful to look into the effects of the external forcing on the spanwise position of the center of the vortex $x_V^*$. Figure \ref{fig:xV_Sth} shows that, with the exception of the baseline flow and of few controlled flows which are assimilable to it (e.g. $St_A \approx 1$), $x_V^*$ is relatively stable and similar to $x_{low}^* = 1/tan\left(\alpha\right)$, which is the position of the lower edge of the ramp. This suggests that $P_b$, that is the wall-pressure at the base of the ramp (see Figure \ref{fig:vortex}), might be related to the pressure field induced by the vortex, in particular at its center.
For simplicity, let us put $x_V^* \approx x_{low}^*$. Then, it will be:
\begin{equation}
C_{p,b} \approx C_{p,V} + \int_{-1-y_V^*}^{-1} \frac{\partial C_p}{\partial y^*} \mathrm{d}y^*.
\label{eq:Cpb_1}
\end{equation}
The non-dimensional vertical pressure gradient $\partial C_p/\partial y^*$ can be computed with the RANS equation along the wall-normal direction, as:
\begin{equation}
\frac{\partial C_p}{\partial y^*} \approx -\frac{2h}{U_\infty^2}\left(U\frac{\partial V}{\partial x} + V\frac{\partial V}{\partial y} - \frac{\partial \langle u^\prime v^\prime \rangle}{\partial x} - \frac{\partial \langle v^{\prime 2} \rangle}{\partial y}\right),
\label{eq:dCp_dy}
\end{equation}
where the symbol $\langle\bullet\rangle$ indicates ensemble averaging.
It is practical to estimate the order of magnitude of $\partial C_p/\partial y^*$ with some dimensional analysis.
An interesting starting point is given by the relationship $b^* \sim L_R^*$ (Eq. \ref{eq:yV_LR}), which has implications on the scaling of velocities within the recirculation region. Indeed, since the backflow must remain constant through all sections of the recirculation region, the following should be verified:
\begin{equation}
-b^* U_R^* \approx \dot{m}_R^* \approx v_{e,R}^*\frac{L_R^*}{2}.
\label{eq:mass_balance_vortex}
\end{equation}
In this expression, $U_R^*<0$ is a characteristic streamwise velocity scale within the recirculation region and $v_{e,R}^*$ is the mean entrainment velocity along the mean separation line. We show in Appendix that, if $L_R^* > 2.5$, $v_{e,R}^*$ is approximately independent of $St_A$ and $v_{e,R}^* = \mathcal{O}\left(10^{-2}\right)$. Of course, Eq. \ref{eq:yV_LR} implies that $U_R \sim U_\infty$ as long as $v_{e,R}^* \approx const.$ , that is, if the recirculation region is not too small, the velocity scale within the recirculation region is similar to the one of the baseline flow, regardless to the working point of the synthetic jets.
If this is so, we can tentatively rely on previous works on natural separated flows to assess the order of magnitude of each term of Eq. \ref{eq:dCp_dy}.
In particular, results reported by \citet{le1997}, \citet{dandois07} and \citet{stella2017} (among others) suggest that $U_R^* = \mathcal{O}\left(10^{-1}\right)$ within the recirculation region, and that the turbulent terms will tend to cancel each other out. Previous sections allow us to also assume that the characteristic horizontal and vertical length scales of the recirculation region will both depend on $L_R^*$. In addition, it seems possible to consider that $V^* \sim v_{e,R}^*$ (for this approximation, the reader is referred to Eq. \ref{eq:ve_Uinf} and \ref{eq:exchangeLenght} in appendix \ref{sec:entrRates}). With these dimensional considerations in mind, Eq. \ref{eq:dCp_dy} reduces to:
\begin{equation}
\frac{\partial C_p}{\partial y^*} \approx - U^*\,\frac{\partial V^*}{\partial x^*} \sim -L_R^{* -1}\mathcal{O}\left(10^{-3}\right).
\label{eq:dCp_dy_2}
\end{equation}
Since $L_R^* = \mathcal{O}(1)$, this leads to:
\begin{equation}
C_{p,b} \approx C_{p,v} \sim \dot{m}_R^*
\label{eq:Cpb_2}
\end{equation}
The evolution of $C_{p,b}$ with respect to $\dot{m}_R^*$ is presented in Figure \ref{fig:Cp_mR}, showing relatively good agreement with the linear trend predicted by Eq. \ref{eq:Cpb_2}. We stress the practical interest of this result: the base pressure $C_{p,b}$, obtainable with a single flush-mounted pressure tap, appears to be a reliable observer of $\dot{m}_R^*$. Since the main characteristic properties of the vortex model, including $L_R^*$, were shown to evolve approximately monotonically with $\dot{m}_R^*$, $C_{p,b}$ might allow to simply reconstruct many fundamental aspects of the large-scale mean topology of separating/reattaching flows, without the need of expensive and unpractical sensing systems.

\section{Conclusions}\label{sec:concluVortex}
In this study we have proposed an original model of mean separating/reattaching flows, which puts the backflow $\dot{m}_R^*$ at the core of their functionning.
Our model represents the mean flow as a large spanwise vortex, covering the entire recirculation region and most of the separated shear layer.
The vortex is defined by two characteristic parameters, its scale $L_R^*$ and its circulation $\Gamma_V^*$. 
To test the relevance of the model, we have analysed the relationship between $\dot{m}_R^*$ and the parameters of the vortex on a set of 21 different mean separated flows, each obtained by forcing a single separated ramp flow at a different actuation frequency. 
The external forcing, provided by a rack of synthetic jets, significantly changes the mean topology of the flow, but it does not impact its global $L_R^*$ scaling.

As a first step, we have estimated $\dot{m}_R^*$ from PIV data, showing that its intensity is approximately inversely correlated to the characteristic jet velocity $U_{JP}^*$.
We have also demonstated that the circulation induced by the vortex $\Gamma_V^*$ scales as $\left(\dot{m}_R^*\right)^2$: as such, $\dot{m}_R^*$ can be assimilated to the amount of mass that is entrained in rotation by the circulation of the vortex.
Since in our model the vortex dominates the separated flow, this result immediately highlights $\dot{m}_R^*$ as one of the key variables driving mean separating/reattaching flows.
On this basis, we have then analysed the correlation between $\dot{m}_R^*$ and $L_R^*$, showing that it can be satisfactory approximated with a linear model. This finding confirms and extends previous studies on natural and forced separating/reattaching flows, suggesting that $\dot{m}_R^*$ is a more appropriate estimator of the variation of $L_R^*$ than other flow or actuation parameters, such as the thickness of the incoming boundary layer or $St_A$. This might pave the way to a universal description of separating/reattaching flows, in particular independent of the characteristics of the actuators.

As a final contribution, we have exploited the vortex model to tackle the problem of estimating $\dot{m}_R^*$ in industrial applications. 
By invoking the Joukovsky theorem, we have proven that $\dot{m}_R^*$ is linearly correlated to the pressure at the center of the vortex, which is itself well approximated by wall-pressure at the base of the ramp.
As such, it appears that a single pressure measurement, simply accessible with a flush-mounted tap, might be sufficient to estimate $\dot{m}_R^*$.
In turn, this might allow us to both reconstruct $L_R^*$ (as well as many large-scale features of the mean flow scaling with it) and close the control loop, by using $\dot{m}_R^*$ as feedback to tune $U_{JP}^*$.

In the light of these findings, our future efforts will pursue two complementary objectives. In the first place, we aim at exploiting mass entrainment to develop new model-based, closed-loop separation control systems. 
In the second place, we would like to extend the vortex model adopted in this study, in the hope of contributing to the development of fast, inexpensive numerical tools to predict the large-scale features of separating/reattaching flows.
\begin{acknowledgments}
This work was supported by the CNRS Groupement De Recherche (GDR) 2502 “Flow Separation Control”, and by the French National Research Agency (ANR) through the Investissements d'Avenir program, under the Labex CAPRYSSES Project (ANR-11-LABX-0006-01). The authors wish to gratefully thank M. St\'{e}phane Loyer (PRISME, Univ. Orl\'{e}ans) for his contribution to wind tunnel measurements.
\end{acknowledgments}

\appendix

\section*{Appendix: a note on entrainment rates}\label{sec:entrRates}
Although not essential to the present discussion of the vortex model, it seems important to complete the analysis of the role of the backflow by briefly investigating the mean \textit{entrainment rate} along the mean separation line. Hereafter, this quantity will be indicated with the symbol $v_{e,R}^*$. 
Interest in the behaviour of $v_{e,R}^*$ is motivated by several previous findings. In particular, \citet{stella2017} shows that $v_{e,R}^*$ plays a crucial role in the development of the separated shear layer (along with the entrainment rate of external fluid from the free flow) and hence in the tuning of $L_R^*$ \cite{adamsJohnstonPart1,barros2016,berk2017}.
In this respect, we believe that, through the study of $v_{e,R}^*$, our experimental database and some of the findings of previous sections can provide new, relevant insight into the interactions between the actuator and the separated flow.

\begin{figure}
  \centering
  \includegraphics[width=\linewidth]{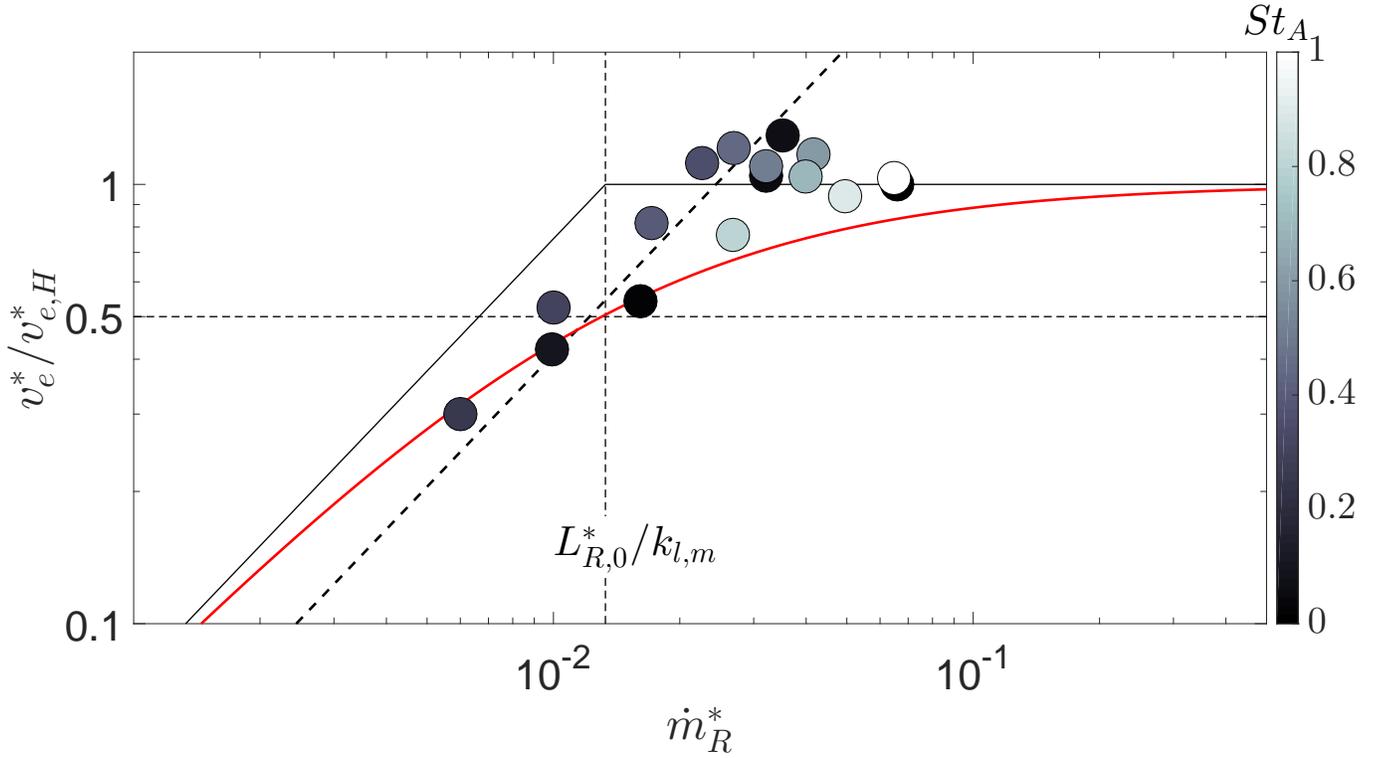}
\caption{Evolution of $v_{e,R}^*$ with $\dot{m}_R^*$. The grayscale indicates the value of $St_A$. The characteristic of the high-pass filter defined by Eq. \ref{eq:ve_Uinf_2} is shown in red online (\protect{\color{red}{\solidrule}}) and its linearised form in fine, black online (\protect\finesolidrule). Fine dashed lines (\protect\finedashedrule) highlight $v_{e,R}^*/v_{e,H}^*=0.5$ and $\dot{m}_R^* = L_R^*/k_{l,m}$. The decreasing trend of $v_{e,R}^*$ for $\dot{m}_R^* < 0.03$ is represented by a thick, dashed line (\protect\dashedrule), which was obtained by fitting a model in the form $v_{e,R}^*/v_{e,H}^*=k_{filt}\,\dot{m}_R^*$, where $k_{filt}$ is a constant, onto available data. The intersection of this linear fit with $v_{e,R}^*/v_{e,H}^*=0.5$ gives $\dot{m}_{R,cut}^* \approx 0.0123$.}
\label{fig:ve_mR}
\end{figure}

According to \citet{stella2017}, $v_{e,R}^*$ can be computed as:
\begin{equation}
v_{e,R}^* = \frac{v_{e,R}}{U_\infty}\approx \frac{\dot{m}_R^*}{S_R^*/2}.
\label{eq:ve_Uinf}
\end{equation}
In this expression, $v_{e,R}$ is a mean entrainment velocity computed along the mean separation line, say on $x/h \in \left(L_R/2, L_R\right)$, and hence, in first approximation:
\begin{equation}
S_R^*/2 = \frac{1}{h}\int_{s_{V}}^{s_1} ds \sim L_R^*,
\label{eq:exchangeLenght}
\end{equation}
where $s_V$ and $s_1$ are the values of $s$ corresponding to $x_V$ and $x_R$, respectively. 
Figure \ref{fig:ve_mR} presents values of $v_{e,R}^*$ yielded by Eq. \ref{eq:ve_Uinf}, for all retained values of $St_A$ (see Sec. \ref{sec:backflowComp}). Eq. \ref{eq:ve_Uinf} suggests that $v_{e,R}^*$ can be interpreted as a measure of efficiency of mass exchange through the mean separation line. Then, $v_{e,R}^*$ is plotted in function of $\dot{m}_R^*$.
Results reported by \citet{stella2017} show that in a unperturbed flow $v_{e,R}^*$ is insensitive to changes in the structure of the separated shear layer, caused by a relatively wide variation of $\Rey_\theta$ in the incoming boundary layer.
Quite surprisingly, a similar behaviour is also observed in Figure \ref{fig:ve_mR}, for $\dot{m}_R^* > 0.03$. On this domain, indeed, it is $v_{e,R}^* \approx v_{e,H}^*\approx 0.021$, which is in very good agreement with entrainment rates observed by \citet{stella2017} ($v_{e,R}^* \approx 0.024 \pm 0.002$). 
Since $\dot{m}_R^*$ is linearly correlated to $L_R^*$ (see Figure \ref{fig:LR_mR}), it appears that for large recirculation regions the efficiency of mass exchanges through the mean separation line is as insensitive to actuation as to $\Rey_\theta$.
In addition, $v_{e,H}^*$ does not seem to be much affected by the value of the parameter $\delta_e/h$ either \cite{adamsJohnstonPart1}, as suggested by comparing the present experiment ($\delta_e/h \approx 0.3$) with \citet{stella2017} ($\delta_e/h \approx 0.9$). 
This being so, it seems likely that $v_{e,H}^*$ might rather depend on factors that were kept similar across experiments, in particular geometric characteristics such as the profile of the ramp, or its expansion ratio (approximately \num{1.1} in both cases).

For $\dot{m}_R^* < 0.03$, Figure \ref{fig:ve_mR} shows that $v_{e,R}^*$ decreases linearly, approximately of a factor 10 per every decade of $\dot{m}_R^*$. Since $\dot{m}_R^*$ is also linearly correlated to $L_R^*$, Eq. \ref{eq:LR_mR} implies that the jets have an effect on $v_{e,R}^*$ only if $L_R^*<2.5$. Such small recirculation regions are attained for $St_A$ approximately within $\left(0.05,0.5\right)$, which broadly corresponds to those actuation frequencies for which $U_{JP}^* \approx U_{JP,max}^*$.
According to Figure \ref{fig:LR_StA}, however, about \SI{75}{\percent} of the variation of $L_R^*$ is achieved for $St_A<0.05$ or $St_A>0.5$. 
In spite of the high values of $U_{JP}^*$, then, the synthetic jets become less effective at modifying $L_R^*$ when they also act on $v_{e,R}^*$.

In this regard, it seems worth pointing out that the trend of $v_{e,R}^*$ shown in Figure \ref{fig:ve_mR} conceptually reminds the transfer function of a high-pass filter, in the form:
\begin{equation}
v_{e,R}^* \approx v_{e,H}^*\frac{\dot{m}_R^*/\dot{m}_{R,cut}^*}{1 + \dot{m}_R^*/\dot{m}_{R,cut}^*}.
\label{eq:transFunct}
\end{equation}
In this expression, $\dot{m}_{R,cut}^*$ is a cutoff value, that can be estimated as the value of $\dot{m}_R^*$ for which $v_{e,R}^* \approx v_{e,H}^*/2$. For the present experiment, available data yield $\dot{m}_{R,cut}^* \approx 0.0123$.
Let us now take a look at Eq. \ref{eq:ve_Uinf} once again. By considering $S_R^* \approx L_R^*$ and plugging in Eq. \ref{eq:LR_mR}, simple manipulations lead to:
\begin{equation}
v_{e,R}^* \approx 2 \frac{\dot{m}_R^*}{L_R^*} \approx \frac{2}{k_{m,L}} \,\frac{\dot{m}_R^*/\left(L_{R,0}^*/k_{m,L}\right)}{1 + \dot{m}_R^*/\left(L_{R,0}^*/k_{m,L}\right)}.
\label{eq:ve_Uinf_2}
\end{equation}
Comparing this latter expression with Eq. \ref{eq:transFunct} gives $\dot{m}_{R,cut}^* = L_{R,0}^*/k_{m,L} \approx 0.0133$ and $v_{e,H}^* = 2/k_{m,L} \approx 0.033$, which are both pleasingly close to measured values, in particular in the case of $\dot{m}_{R,cut}^*$.
These findings have several important implications.
Firstly, the evolution of $v_{e,R}^*$ with $\dot{m}_R^*$ seems to support the linear correlation between $\dot{m}_R^*$ and $L_R^*$ proposed in Eq. \ref{eq:LR_mR}, including for what concerns the existence of a non null, minimum recirculation length $L_{R,0}^*$.
Secondly, Eq. \ref{eq:ve_Uinf} and Eq. \ref{eq:ve_Uinf_2} might provide interesting insight into the physical significance of both parameters of Eq. \ref{eq:LR_mR}. In particular, the proportionality constant $k_{m,L}$ does not depend on actuation parameters, nor on properties determined by the incoming boundary layer, such as $\Rey_\theta$ or $\delta_e/h$, because $k_{m,L} \sim \left(v_{e,H}^*\right)^{-1}$ and $v_{e,H}^*\approx 0.02$ in both the present experiment and the unperturbed flow analysed in \citet{stella2017}. Since the two works share the same ramp profile and have similar expansion ratios, it seems more likely that $k_{m,L}$ mostly depends on geometry. As for what concerns $L_{R,0}^*$, it is tempting to relate its value to the frequency response of the actuators. Indeed, Eq. \ref{eq:LR_mR} yields:
\begin{equation}
L_{R,cut}^* = L_R^*\left(\dot{m}_{R,cut}^*\right) = 2 L_{R,0}^* \approx 1.6 \approx L_{R,min}^*,
\end{equation}
and hence $L_{R,0}^* \sim L_{R,min}^*$.
Now, curves reported in Figure \ref{fig:Freq_resp} and Figure \ref{fig:LR_StA} suggest that $L_R^* \rightarrow L_{R,min}^*$ is broadly correlated to $U_{JP}^* \rightarrow U_{JP,max}^*$, i.e. that $L_{R,min}^*$ might be mainly determined by the saturation of the synthetic jets. Then, although at present no element allows to definitely discard an additional dependency on geometry, $L_{R,0}^*$ might be rather influenced by properties of the actuators.
Finally, it seems important to remind that, at least to a certain extent, $U_{JP}^*$ can be considered anticorrelated to $\dot{m}_R^*$ (see Figure \ref{fig:Freq_resp}). This means that it should be possible to transpose Eq. \ref{eq:ve_Uinf_2} into a \textit{low-pass} filter that applies to the velocity of the synthetic jet. In other words, the control action becomes ineffective at reducing $L_R^*$ if $U_{JP}^*$ exceeds a value corresponding to $\dot{m}_{R,cut}^*$.  
Most importantly, for a given actuator this threshold seems to mainly depend on the geometry of the ramp. 

\FloatBarrier
\bibliography{2018_SJMK}

\begin{thebibliography}{36}%
\makeatletter
\providecommand \@ifxundefined [1]{%
 \@ifx{#1\undefined}
}%
\providecommand \@ifnum [1]{%
 \ifnum #1\expandafter \@firstoftwo
 \else \expandafter \@secondoftwo
 \fi
}%
\providecommand \@ifx [1]{%
 \ifx #1\expandafter \@firstoftwo
 \else \expandafter \@secondoftwo
 \fi
}%
\providecommand \natexlab [1]{#1}%
\providecommand \enquote  [1]{``#1''}%
\providecommand \bibnamefont  [1]{#1}%
\providecommand \bibfnamefont [1]{#1}%
\providecommand \citenamefont [1]{#1}%
\providecommand \href@noop [0]{\@secondoftwo}%
\providecommand \href [0]{\begingroup \@sanitize@url \@href}%
\providecommand \@href[1]{\@@startlink{#1}\@@href}%
\providecommand \@@href[1]{\endgroup#1\@@endlink}%
\providecommand \@sanitize@url [0]{\catcode `\\12\catcode `\$12\catcode
  `\&12\catcode `\#12\catcode `\^12\catcode `\_12\catcode `\%12\relax}%
\providecommand \@@startlink[1]{}%
\providecommand \@@endlink[0]{}%
\providecommand \url  [0]{\begingroup\@sanitize@url \@url }%
\providecommand \@url [1]{\endgroup\@href {#1}{\urlprefix }}%
\providecommand \urlprefix  [0]{URL }%
\providecommand \Eprint [0]{\href }%
\providecommand \doibase [0]{http://dx.doi.org/}%
\providecommand \selectlanguage [0]{\@gobble}%
\providecommand \bibinfo  [0]{\@secondoftwo}%
\providecommand \bibfield  [0]{\@secondoftwo}%
\providecommand \translation [1]{[#1]}%
\providecommand \BibitemOpen [0]{}%
\providecommand \bibitemStop [0]{}%
\providecommand \bibitemNoStop [0]{.\EOS\space}%
\providecommand \EOS [0]{\spacefactor3000\relax}%
\providecommand \BibitemShut  [1]{\csname bibitem#1\endcsname}%
\let\auto@bib@innerbib\@empty
\bibitem [{\citenamefont {Seifert}\ \emph {et~al.}(2015)\citenamefont
  {Seifert}, \citenamefont {Shtendel},\ and\ \citenamefont
  {Dolgopyat}}]{seifert2015}%
  \BibitemOpen
  \bibfield  {author} {\bibinfo {author} {\bibfnamefont {A.}~\bibnamefont
  {Seifert}}, \bibinfo {author} {\bibfnamefont {T.}~\bibnamefont {Shtendel}}, \
  and\ \bibinfo {author} {\bibfnamefont {D.}~\bibnamefont {Dolgopyat}},\
  }\bibfield  {title} {\enquote {\bibinfo {title} {From lab to full scale
  {Active} {Flow} {Control} drag reduction: {How} to bridge the gap?}}\
  }\href@noop {} {\bibfield  {journal} {\bibinfo  {journal} {J. Wind Eng. Ind.
  Aerodyn.}\ }\textbf {\bibinfo {volume} {147}},\ \bibinfo {pages} {262--272}
  (\bibinfo {year} {2015})}\BibitemShut {NoStop}%
\bibitem [{\citenamefont {Armaly}\ \emph {et~al.}(1983)\citenamefont {Armaly},
  \citenamefont {Durst}, \citenamefont {Pereira},\ and\ \citenamefont
  {Sch{\"o}nung}}]{armaly1983}%
  \BibitemOpen
  \bibfield  {author} {\bibinfo {author} {\bibfnamefont {B.~F.}\ \bibnamefont
  {Armaly}}, \bibinfo {author} {\bibfnamefont {F.}~\bibnamefont {Durst}},
  \bibinfo {author} {\bibfnamefont {J.~C.~F.}\ \bibnamefont {Pereira}}, \ and\
  \bibinfo {author} {\bibfnamefont {B.}~\bibnamefont {Sch{\"o}nung}},\
  }\bibfield  {title} {\enquote {\bibinfo {title} {Experimental and theoretical
  investigation of backward-facing step flow},}\ }\href@noop {} {\bibfield
  {journal} {\bibinfo  {journal} {J. Fluid Mech.}\ }\textbf {\bibinfo {volume}
  {127}},\ \bibinfo {pages} {473--496} (\bibinfo {year} {1983})}\BibitemShut
  {NoStop}%
\bibitem [{\citenamefont {Simpson}(1989)}]{simpson1989}%
  \BibitemOpen
  \bibfield  {author} {\bibinfo {author} {\bibfnamefont {R.~L.}\ \bibnamefont
  {Simpson}},\ }\bibfield  {title} {\enquote {\bibinfo {title} {Turbulent
  boundary-layer separation},}\ }\href@noop {} {\bibfield  {journal} {\bibinfo
  {journal} {Annu. Rev. Fluid Mech.}\ }\textbf {\bibinfo {volume} {21}},\
  \bibinfo {pages} {205--232} (\bibinfo {year} {1989})}\BibitemShut {NoStop}%
\bibitem [{\citenamefont {Dandois}\ \emph {et~al.}(2007)\citenamefont
  {Dandois}, \citenamefont {Garnier},\ and\ \citenamefont
  {Sagaut}}]{dandois07}%
  \BibitemOpen
  \bibfield  {author} {\bibinfo {author} {\bibfnamefont {J.}~\bibnamefont
  {Dandois}}, \bibinfo {author} {\bibfnamefont {E.}~\bibnamefont {Garnier}}, \
  and\ \bibinfo {author} {\bibfnamefont {P.}~\bibnamefont {Sagaut}},\
  }\bibfield  {title} {\enquote {\bibinfo {title} {Numerical simulation of
  active separation control by a synthetic jet},}\ }\href@noop {} {\bibfield
  {journal} {\bibinfo  {journal} {J. Fluid Mech.}\ }\textbf {\bibinfo {volume}
  {574}},\ \bibinfo {pages} {25--58} (\bibinfo {year} {2007})}\BibitemShut
  {NoStop}%
\bibitem [{\citenamefont {Stella}\ \emph {et~al.}(2017)\citenamefont {Stella},
  \citenamefont {Mazellier},\ and\ \citenamefont {Kourta}}]{stella2017}%
  \BibitemOpen
  \bibfield  {author} {\bibinfo {author} {\bibfnamefont {F.}~\bibnamefont
  {Stella}}, \bibinfo {author} {\bibfnamefont {N.}~\bibnamefont {Mazellier}}, \
  and\ \bibinfo {author} {\bibfnamefont {A.}~\bibnamefont {Kourta}},\
  }\bibfield  {title} {\enquote {\bibinfo {title} {Scaling of separated shear
  layers: an investigation of mass entrainment},}\ }\href@noop {} {\bibfield
  {journal} {\bibinfo  {journal} {J. Fluid Mech.}\ }\textbf {\bibinfo {volume}
  {826}},\ \bibinfo {pages} {851--887} (\bibinfo {year} {2017})}\BibitemShut
  {NoStop}%
\bibitem [{\citenamefont {Debien}\ \emph {et~al.}(2014)\citenamefont {Debien},
  \citenamefont {Aubrun}, \citenamefont {Mazellier},\ and\ \citenamefont
  {Kourta}}]{debien14}%
  \BibitemOpen
  \bibfield  {author} {\bibinfo {author} {\bibfnamefont {A.}~\bibnamefont
  {Debien}}, \bibinfo {author} {\bibfnamefont {S.}~\bibnamefont {Aubrun}},
  \bibinfo {author} {\bibfnamefont {N.}~\bibnamefont {Mazellier}}, \ and\
  \bibinfo {author} {\bibfnamefont {A.}~\bibnamefont {Kourta}},\ }\bibfield
  {title} {\enquote {\bibinfo {title} {Salient and smooth edge ramps inducing
  turbulent boundary layer separation: {Flow} characterization for control
  perspective},}\ }\href@noop {} {\bibfield  {journal} {\bibinfo  {journal}
  {Comptes Rendus Mécanique: Flow separation control}\ }\textbf {\bibinfo
  {volume} {342}},\ \bibinfo {pages} {356--362} (\bibinfo {year}
  {2014})}\BibitemShut {NoStop}%
\bibitem [{\citenamefont {Kourta}\ \emph {et~al.}(2015)\citenamefont {Kourta},
  \citenamefont {Thacker},\ and\ \citenamefont {Joussot}}]{kourta15}%
  \BibitemOpen
  \bibfield  {author} {\bibinfo {author} {\bibfnamefont {A.}~\bibnamefont
  {Kourta}}, \bibinfo {author} {\bibfnamefont {A.}~\bibnamefont {Thacker}}, \
  and\ \bibinfo {author} {\bibfnamefont {R.}~\bibnamefont {Joussot}},\
  }\bibfield  {title} {\enquote {\bibinfo {title} {Analysis and
  characterization of ramp flow separation},}\ }\href@noop {} {\bibfield
  {journal} {\bibinfo  {journal} {Exp. Fluids}\ }\textbf {\bibinfo {volume}
  {56}},\ \bibinfo {pages} {1--14} (\bibinfo {year} {2015})}\BibitemShut
  {NoStop}%
\bibitem [{\citenamefont {Brown}\ and\ \citenamefont
  {Roshko}(1974)}]{brownRoshko74}%
  \BibitemOpen
  \bibfield  {author} {\bibinfo {author} {\bibfnamefont {G.~L.}\ \bibnamefont
  {Brown}}\ and\ \bibinfo {author} {\bibfnamefont {A.}~\bibnamefont {Roshko}},\
  }\bibfield  {title} {\enquote {\bibinfo {title} {On density effects and large
  structure in turbulent mixing layers},}\ }\href@noop {} {\bibfield  {journal}
  {\bibinfo  {journal} {J. Fluid Mech.}\ }\textbf {\bibinfo {volume} {64}},\
  \bibinfo {pages} {775--816} (\bibinfo {year} {1974})}\BibitemShut {NoStop}%
\bibitem [{\citenamefont {Le}\ \emph {et~al.}(1997)\citenamefont {Le},
  \citenamefont {Moin},\ and\ \citenamefont {Kim}}]{le1997}%
  \BibitemOpen
  \bibfield  {author} {\bibinfo {author} {\bibfnamefont {H.}~\bibnamefont
  {Le}}, \bibinfo {author} {\bibfnamefont {P.}~\bibnamefont {Moin}}, \ and\
  \bibinfo {author} {\bibfnamefont {J.}~\bibnamefont {Kim}},\ }\bibfield
  {title} {\enquote {\bibinfo {title} {Direct numerical simulation of turbulent
  flow over a backward-facing step},}\ }\href@noop {} {\bibfield  {journal}
  {\bibinfo  {journal} {J. Fluid Mech.}\ }\textbf {\bibinfo {volume} {330}},\
  \bibinfo {pages} {349--374} (\bibinfo {year} {1997})}\BibitemShut {NoStop}%
\bibitem [{\citenamefont {Barros}\ \emph {et~al.}(2016)\citenamefont {Barros},
  \citenamefont {Borée}, \citenamefont {Noack}, \citenamefont {Spohn},\ and\
  \citenamefont {Ruiz}}]{barros2016}%
  \BibitemOpen
  \bibfield  {author} {\bibinfo {author} {\bibfnamefont {D.}~\bibnamefont
  {Barros}}, \bibinfo {author} {\bibfnamefont {J.}~\bibnamefont {Borée}},
  \bibinfo {author} {\bibfnamefont {B.~R.}\ \bibnamefont {Noack}}, \bibinfo
  {author} {\bibfnamefont {A}~\bibnamefont {Spohn}}, \ and\ \bibinfo {author}
  {\bibfnamefont {T.}~\bibnamefont {Ruiz}},\ }\bibfield  {title} {\enquote
  {\bibinfo {title} {Bluff body drag manipulation using pulsed jets and
  {Coanda} effect},}\ }\href@noop {} {\bibfield  {journal} {\bibinfo  {journal}
  {J. Fluid Mech.}\ }\textbf {\bibinfo {volume} {805}},\ \bibinfo {pages}
  {422--459} (\bibinfo {year} {2016})}\BibitemShut {NoStop}%
\bibitem [{\citenamefont {Pujals}\ \emph {et~al.}(2010)\citenamefont {Pujals},
  \citenamefont {Depardon},\ and\ \citenamefont {Cossu}}]{pujals2010}%
  \BibitemOpen
  \bibfield  {author} {\bibinfo {author} {\bibfnamefont {G.}~\bibnamefont
  {Pujals}}, \bibinfo {author} {\bibfnamefont {S.}~\bibnamefont {Depardon}}, \
  and\ \bibinfo {author} {\bibfnamefont {C.}~\bibnamefont {Cossu}},\ }\bibfield
   {title} {\enquote {\bibinfo {title} {Drag reduction of a 3d bluff body using
  coherent streamwise streaks},}\ }\href@noop {} {\bibfield  {journal}
  {\bibinfo  {journal} {Exp Fluids}\ }\textbf {\bibinfo {volume} {49}},\
  \bibinfo {pages} {1085--1094} (\bibinfo {year} {2010})}\BibitemShut {NoStop}%
\bibitem [{\citenamefont {Rouméas}\ \emph {et~al.}(2009)\citenamefont
  {Rouméas}, \citenamefont {Gilliéron},\ and\ \citenamefont
  {Kourta}}]{roumeas2009}%
  \BibitemOpen
  \bibfield  {author} {\bibinfo {author} {\bibfnamefont {M.}~\bibnamefont
  {Rouméas}}, \bibinfo {author} {\bibfnamefont {P.}~\bibnamefont
  {Gilliéron}}, \ and\ \bibinfo {author} {\bibfnamefont {A.}~\bibnamefont
  {Kourta}},\ }\bibfield  {title} {\enquote {\bibinfo {title} {Drag reduction
  by flow separation control on a car after body},}\ }\href@noop {} {\bibfield
  {journal} {\bibinfo  {journal} {Int. J. Numer. Methods Fluids}\ }\textbf
  {\bibinfo {volume} {60}},\ \bibinfo {pages} {1222--1240} (\bibinfo {year}
  {2009})}\BibitemShut {NoStop}%
\bibitem [{\citenamefont {Donovan}\ \emph {et~al.}(1997)\citenamefont
  {Donovan}, \citenamefont {Kral},\ and\ \citenamefont {Cary}}]{donovan1997}%
  \BibitemOpen
  \bibfield  {author} {\bibinfo {author} {\bibfnamefont {J.}~\bibnamefont
  {Donovan}}, \bibinfo {author} {\bibfnamefont {L.}~\bibnamefont {Kral}}, \
  and\ \bibinfo {author} {\bibfnamefont {A.}~\bibnamefont {Cary}},\ }\bibfield
  {title} {\enquote {\bibinfo {title} {Active flow control applied to an
  airfoil},}\ }in\ \href@noop {} {\emph {\bibinfo {booktitle} {36th {AIAA}
  {Aerospace} {Sciences} {Meeting} and {Exhibit}}}}\ (\bibinfo  {publisher}
  {AIAA},\ \bibinfo {address} {Reno, NV,U.S.A.},\ \bibinfo {year}
  {1997})\BibitemShut {NoStop}%
\bibitem [{\citenamefont {Joseph}\ \emph {et~al.}(2012)\citenamefont {Joseph},
  \citenamefont {Amandolèse},\ and\ \citenamefont {Aider}}]{joseph2012}%
  \BibitemOpen
  \bibfield  {author} {\bibinfo {author} {\bibfnamefont {P.}~\bibnamefont
  {Joseph}}, \bibinfo {author} {\bibfnamefont {X.}~\bibnamefont {Amandolèse}},
  \ and\ \bibinfo {author} {\bibfnamefont {J.-L.}\ \bibnamefont {Aider}},\
  }\bibfield  {title} {\enquote {\bibinfo {title} {Drag reduction on the
  25${}^\circ$ slant angle {Ahmed} reference body using pulsed jets},}\
  }\href@noop {} {\bibfield  {journal} {\bibinfo  {journal} {Exp. Fluids}\
  }\textbf {\bibinfo {volume} {52}},\ \bibinfo {pages} {1169--1185} (\bibinfo
  {year} {2012})}\BibitemShut {NoStop}%
\bibitem [{\citenamefont {Thomas}\ \emph {et~al.}(2008)\citenamefont {Thomas},
  \citenamefont {Kozlov},\ and\ \citenamefont {Corke}}]{thomas2008}%
  \BibitemOpen
  \bibfield  {author} {\bibinfo {author} {\bibfnamefont {F.~O.}\ \bibnamefont
  {Thomas}}, \bibinfo {author} {\bibfnamefont {A.}~\bibnamefont {Kozlov}}, \
  and\ \bibinfo {author} {\bibfnamefont {T.~C.}\ \bibnamefont {Corke}},\
  }\bibfield  {title} {\enquote {\bibinfo {title} {Plasma {Actuators} for
  {Cylinder} {Flow} {Control} and {Noise} {Reduction}},}\ }\href@noop {}
  {\bibfield  {journal} {\bibinfo  {journal} {AIAA J.}\ }\textbf {\bibinfo
  {volume} {46}},\ \bibinfo {pages} {1921--1931} (\bibinfo {year}
  {2008})}\BibitemShut {NoStop}%
\bibitem [{\citenamefont {Kourta}\ and\ \citenamefont
  {Leclerc}(2013)}]{kourta2013}%
  \BibitemOpen
  \bibfield  {author} {\bibinfo {author} {\bibfnamefont {A.}~\bibnamefont
  {Kourta}}\ and\ \bibinfo {author} {\bibfnamefont {C.}~\bibnamefont
  {Leclerc}},\ }\bibfield  {title} {\enquote {\bibinfo {title}
  {Characterization of synthetic jet actuation with application to {Ahmed} body
  wake},}\ }\href@noop {} {\bibfield  {journal} {\bibinfo  {journal} {Sensors
  Actuators A: Phys.}\ }\textbf {\bibinfo {volume} {192}},\ \bibinfo {pages}
  {13--26} (\bibinfo {year} {2013})}\BibitemShut {NoStop}%
\bibitem [{\citenamefont {Shimizu}\ \emph {et~al.}(1993)\citenamefont
  {Shimizu}, \citenamefont {Kiya}, \citenamefont {Mochizuki},\ and\
  \citenamefont {Ido}}]{shimizu1993}%
  \BibitemOpen
  \bibfield  {author} {\bibinfo {author} {\bibfnamefont {M.}~\bibnamefont
  {Shimizu}}, \bibinfo {author} {\bibfnamefont {M.}~\bibnamefont {Kiya}},
  \bibinfo {author} {\bibfnamefont {O.}~\bibnamefont {Mochizuki}}, \ and\
  \bibinfo {author} {\bibfnamefont {Y.}~\bibnamefont {Ido}},\ }\bibfield
  {title} {\enquote {\bibinfo {title} {Response of an axisymmetric separation
  bubble to sinusoidal forcing - {Effects} of forcing frequency, forcing level
  and {Reynolds} number},}\ }\href@noop {} {\bibfield  {journal} {\bibinfo
  {journal} {JSME Trans.}\ }\textbf {\bibinfo {volume} {59}},\ \bibinfo {pages}
  {721--727} (\bibinfo {year} {1993})}\BibitemShut {NoStop}%
\bibitem [{\citenamefont {Sigurdson}(1995)}]{sigurdson1995}%
  \BibitemOpen
  \bibfield  {author} {\bibinfo {author} {\bibfnamefont {L.~W.}\ \bibnamefont
  {Sigurdson}},\ }\bibfield  {title} {\enquote {\bibinfo {title} {The structure
  and control of a turbulent reattaching flow},}\ }\href@noop {} {\bibfield
  {journal} {\bibinfo  {journal} {J. Fluid Mech.}\ }\textbf {\bibinfo {volume}
  {298}},\ \bibinfo {pages} {139--165} (\bibinfo {year} {1995})}\BibitemShut
  {NoStop}%
\bibitem [{\citenamefont {Chun}\ and\ \citenamefont {Sung}(1996)}]{chun1996}%
  \BibitemOpen
  \bibfield  {author} {\bibinfo {author} {\bibfnamefont {K.~B.}\ \bibnamefont
  {Chun}}\ and\ \bibinfo {author} {\bibfnamefont {H.~J.}\ \bibnamefont
  {Sung}},\ }\bibfield  {title} {\enquote {\bibinfo {title} {Control of
  turbulent separated flow over a backward-facing step by local forcing},}\
  }\href@noop {} {\bibfield  {journal} {\bibinfo  {journal} {Exp. Fluids}\
  }\textbf {\bibinfo {volume} {21}},\ \bibinfo {pages} {417--426} (\bibinfo
  {year} {1996})}\BibitemShut {NoStop}%
\bibitem [{\citenamefont {Glezer}\ \emph {et~al.}(2005)\citenamefont {Glezer},
  \citenamefont {Amitay},\ and\ \citenamefont {Honohan}}]{glezer2005}%
  \BibitemOpen
  \bibfield  {author} {\bibinfo {author} {\bibfnamefont {A.}~\bibnamefont
  {Glezer}}, \bibinfo {author} {\bibfnamefont {M.}~\bibnamefont {Amitay}}, \
  and\ \bibinfo {author} {\bibfnamefont {A.~M.}\ \bibnamefont {Honohan}},\
  }\bibfield  {title} {\enquote {\bibinfo {title} {Aspects of {Low}- and
  {High}-{Frequency} {Actuation} for {Aerodynamic} {Flow} {Control}},}\ }\href
  {\doibase 10.2514/1.7411} {\bibfield  {journal} {\bibinfo  {journal} {AIAA
  J.}\ }\textbf {\bibinfo {volume} {43}},\ \bibinfo {pages} {1501--1511}
  (\bibinfo {year} {2005})}\BibitemShut {NoStop}%
\bibitem [{\citenamefont {Parezanović}\ \emph {et~al.}(2015)\citenamefont
  {Parezanović}, \citenamefont {Laurentie}, \citenamefont {Fourment},
  \citenamefont {Delville}, \citenamefont {Bonnet}, \citenamefont {Spohn},
  \citenamefont {Duriez}, \citenamefont {Cordier}, \citenamefont {Noack},
  \citenamefont {Abel}, \citenamefont {Segond}, \citenamefont {Shaqarin},\ and\
  \citenamefont {Brunton}}]{parezanovic2015}%
  \BibitemOpen
  \bibfield  {author} {\bibinfo {author} {\bibfnamefont {V.}~\bibnamefont
  {Parezanović}}, \bibinfo {author} {\bibfnamefont {J.-C.}\ \bibnamefont
  {Laurentie}}, \bibinfo {author} {\bibfnamefont {C.}~\bibnamefont {Fourment}},
  \bibinfo {author} {\bibfnamefont {J.}~\bibnamefont {Delville}}, \bibinfo
  {author} {\bibfnamefont {J.-P.}\ \bibnamefont {Bonnet}}, \bibinfo {author}
  {\bibfnamefont {A.}~\bibnamefont {Spohn}}, \bibinfo {author} {\bibfnamefont
  {T.}~\bibnamefont {Duriez}}, \bibinfo {author} {\bibfnamefont
  {L.}~\bibnamefont {Cordier}}, \bibinfo {author} {\bibfnamefont {B.~R.}\
  \bibnamefont {Noack}}, \bibinfo {author} {\bibfnamefont {M.}~\bibnamefont
  {Abel}}, \bibinfo {author} {\bibfnamefont {M.}~\bibnamefont {Segond}},
  \bibinfo {author} {\bibfnamefont {T.}~\bibnamefont {Shaqarin}}, \ and\
  \bibinfo {author} {\bibfnamefont {S.~L.}\ \bibnamefont {Brunton}},\
  }\bibfield  {title} {\enquote {\bibinfo {title} {Mixing {Layer}
  {Manipulation} {Experiment}},}\ }\href@noop {} {\bibfield  {journal}
  {\bibinfo  {journal} {Flow Turbul. Combust.}\ }\textbf {\bibinfo {volume}
  {94}},\ \bibinfo {pages} {155--173} (\bibinfo {year} {2015})}\BibitemShut
  {NoStop}%
\bibitem [{\citenamefont {Berk}\ \emph {et~al.}(2017)\citenamefont {Berk},
  \citenamefont {Medjnoun},\ and\ \citenamefont
  {Ganapathisubramani}}]{berk2017}%
  \BibitemOpen
  \bibfield  {author} {\bibinfo {author} {\bibfnamefont {T.}~\bibnamefont
  {Berk}}, \bibinfo {author} {\bibfnamefont {T.}~\bibnamefont {Medjnoun}}, \
  and\ \bibinfo {author} {\bibfnamefont {B.}~\bibnamefont
  {Ganapathisubramani}},\ }\bibfield  {title} {\enquote {\bibinfo {title}
  {Entrainment effects in periodic forcing of the flow over a backward-facing
  step},}\ }\href@noop {} {\bibfield  {journal} {\bibinfo  {journal} {Phys.
  Rev. Fluids}\ }\textbf {\bibinfo {volume} {2}},\ \bibinfo {pages} {074605}
  (\bibinfo {year} {2017})}\BibitemShut {NoStop}%
\bibitem [{\citenamefont {Adams}\ and\ \citenamefont
  {Johnston}(1988{\natexlab{a}})}]{adamsJohnstonPart1}%
  \BibitemOpen
  \bibfield  {author} {\bibinfo {author} {\bibfnamefont {E.~W.}\ \bibnamefont
  {Adams}}\ and\ \bibinfo {author} {\bibfnamefont {J.~P.}\ \bibnamefont
  {Johnston}},\ }\bibfield  {title} {\enquote {\bibinfo {title} {Effects of the
  separating shear layer on the reattachment flow structure. {Part} 1:
  {Pressure} and turbulence quantities},}\ }\href@noop {} {\bibfield  {journal}
  {\bibinfo  {journal} {Exp. Fluids}\ }\textbf {\bibinfo {volume} {6}},\
  \bibinfo {pages} {400--408} (\bibinfo {year}
  {1988}{\natexlab{a}})}\BibitemShut {NoStop}%
\bibitem [{\citenamefont {Nadge}\ and\ \citenamefont
  {Govardhan}(2014)}]{nadge14}%
  \BibitemOpen
  \bibfield  {author} {\bibinfo {author} {\bibfnamefont {P.~M.}\ \bibnamefont
  {Nadge}}\ and\ \bibinfo {author} {\bibfnamefont {R.~N.}\ \bibnamefont
  {Govardhan}},\ }\bibfield  {title} {\enquote {\bibinfo {title} {High
  {Reynolds} number flow over a backward-facing step: structure of the mean
  separation bubble},}\ }\href@noop {} {\bibfield  {journal} {\bibinfo
  {journal} {Exp. Fluids}\ }\textbf {\bibinfo {volume} {55}},\ \bibinfo {pages}
  {1--22} (\bibinfo {year} {2014})}\BibitemShut {NoStop}%
\bibitem [{\citenamefont {Durst}\ and\ \citenamefont
  {Tropea}(1981)}]{durst1981}%
  \BibitemOpen
  \bibfield  {author} {\bibinfo {author} {\bibfnamefont {F}~\bibnamefont
  {Durst}}\ and\ \bibinfo {author} {\bibfnamefont {C}~\bibnamefont {Tropea}},\
  }\bibfield  {title} {\enquote {\bibinfo {title} {Turbulent, backward-facing
  step flows in two-dimensional ducts and channels},}\ }in\ \href@noop {}
  {\emph {\bibinfo {booktitle} {Proc. 3rd Int. Symp. On Turbulent Shear
  Flows}}}\ (\bibinfo {year} {1981})\ pp.\ \bibinfo {pages} {9--11}\BibitemShut
  {NoStop}%
\bibitem [{\citenamefont {Ruck}\ and\ \citenamefont
  {Makiola}(1993)}]{ruck1993}%
  \BibitemOpen
  \bibfield  {author} {\bibinfo {author} {\bibfnamefont {B.}~\bibnamefont
  {Ruck}}\ and\ \bibinfo {author} {\bibfnamefont {B.}~\bibnamefont {Makiola}},\
  }\bibfield  {title} {\enquote {\bibinfo {title} {Flow {Separation} over the
  {Inclined} {Step}},}\ }in\ \href@noop {} {\emph {\bibinfo {booktitle}
  {Physics of {Separated} {Flows} — {Numerical}, {Experimental}, and
  {Theoretical} {Aspects}}}},\ \bibinfo {series} {Notes on {Numerical} {Fluid}
  {Mechanics} ({NNFM})}, Vol.~\bibinfo {volume} {40}\ (\bibinfo  {publisher}
  {Vieweg+Teubner Verlag, Wiesbaden},\ \bibinfo {year} {1993})\ pp.\ \bibinfo
  {pages} {47--55}\BibitemShut {NoStop}%
\bibitem [{\citenamefont {Adams}\ and\ \citenamefont
  {Johnston}(1988{\natexlab{b}})}]{adamsJohnstonPart2}%
  \BibitemOpen
  \bibfield  {author} {\bibinfo {author} {\bibfnamefont {E.~W.}\ \bibnamefont
  {Adams}}\ and\ \bibinfo {author} {\bibfnamefont {J.~P.}\ \bibnamefont
  {Johnston}},\ }\bibfield  {title} {\enquote {\bibinfo {title} {Effects of the
  separating shear layer on the reattachment flow structure. {Part} 2:
  {Reattachment} length and wall shear stress},}\ }\href@noop {} {\bibfield
  {journal} {\bibinfo  {journal} {Exp. Fluids}\ }\textbf {\bibinfo {volume}
  {6}},\ \bibinfo {pages} {493--499} (\bibinfo {year}
  {1988}{\natexlab{b}})}\BibitemShut {NoStop}%
\bibitem [{\citenamefont {Chapman}\ \emph {et~al.}(1958)\citenamefont
  {Chapman}, \citenamefont {Kuehn},\ and\ \citenamefont
  {Larson}}]{chapman1958}%
  \BibitemOpen
  \bibfield  {author} {\bibinfo {author} {\bibfnamefont {D.~R.}\ \bibnamefont
  {Chapman}}, \bibinfo {author} {\bibfnamefont {D.~M.}\ \bibnamefont {Kuehn}},
  \ and\ \bibinfo {author} {\bibfnamefont {H.~K.}\ \bibnamefont {Larson}},\
  }\href@noop {} {\emph {\bibinfo {title} {Investigation of {Separated} {Flows}
  in {Supersonic} and {Subsonic} {Streams} with {Emphasis} on the {Effect} of
  {Transition}}}},\ \bibinfo {type} {Technical {Report}}\ \bibinfo {number}
  {TN-1356}\ (\bibinfo  {institution} {NACA},\ \bibinfo {address} {Washington,
  DC},\ \bibinfo {year} {1958})\BibitemShut {NoStop}%
\bibitem [{\citenamefont {Eaton}\ and\ \citenamefont
  {Johnston}(1981)}]{eatonJohnson1981}%
  \BibitemOpen
  \bibfield  {author} {\bibinfo {author} {\bibfnamefont {J.~K.}\ \bibnamefont
  {Eaton}}\ and\ \bibinfo {author} {\bibfnamefont {J.~P.}\ \bibnamefont
  {Johnston}},\ }\bibfield  {title} {\enquote {\bibinfo {title} {A {Review} of
  {Research} on {Subsonic} {Turbulent} {Flow} {Reattachment}},}\ }\href@noop {}
  {\bibfield  {journal} {\bibinfo  {journal} {AIAA J.}\ }\textbf {\bibinfo
  {volume} {19}},\ \bibinfo {pages} {1093--1100} (\bibinfo {year}
  {1981})}\BibitemShut {NoStop}%
\bibitem [{\citenamefont {Batchelor}(2000)}]{batchelor2000}%
  \BibitemOpen
  \bibfield  {author} {\bibinfo {author} {\bibfnamefont {G.~K.}\ \bibnamefont
  {Batchelor}},\ }\href@noop {} {\emph {\bibinfo {title} {An {Introduction} to
  {Fluid} {Dynamics}}}}\ (\bibinfo  {publisher} {Cambridge University Press},\
  \bibinfo {year} {2000})\BibitemShut {NoStop}%
\bibitem [{\citenamefont {Hunt}\ \emph {et~al.}(1988)\citenamefont {Hunt},
  \citenamefont {Wray},\ and\ \citenamefont {Moin}}]{hunt1988}%
  \BibitemOpen
  \bibfield  {author} {\bibinfo {author} {\bibfnamefont {J.~C.~R.}\
  \bibnamefont {Hunt}}, \bibinfo {author} {\bibfnamefont {A.~A.}\ \bibnamefont
  {Wray}}, \ and\ \bibinfo {author} {\bibfnamefont {P.}~\bibnamefont {Moin}},\
  }\href@noop {} {\emph {\bibinfo {title} {Eddies, streams, and convergence
  zones in turbulent flows}}},\ \bibinfo {type} {Tech. Rep.}\ \bibinfo {number}
  {Center for Turbulence Research Report CTR-S88}\ (\bibinfo {year}
  {1988})\BibitemShut {NoStop}%
\bibitem [{\citenamefont {Jeong}\ and\ \citenamefont
  {Hussain}(1995)}]{jeong1995}%
  \BibitemOpen
  \bibfield  {author} {\bibinfo {author} {\bibfnamefont {J.}~\bibnamefont
  {Jeong}}\ and\ \bibinfo {author} {\bibfnamefont {F.}~\bibnamefont
  {Hussain}},\ }\bibfield  {title} {\enquote {\bibinfo {title} {On the
  identification of a vortex},}\ }\href@noop {} {\bibfield  {journal} {\bibinfo
   {journal} {J. Fluid Mech.}\ }\textbf {\bibinfo {volume} {285}},\ \bibinfo
  {pages} {69--94} (\bibinfo {year} {1995})}\BibitemShut {NoStop}%
\bibitem [{\citenamefont {Debien}\ \emph {et~al.}(2016)\citenamefont {Debien},
  \citenamefont {von Krbek}, \citenamefont {Mazellier}, \citenamefont {Duriez},
  \citenamefont {Cordier}, \citenamefont {Noack}, \citenamefont {Abel},\ and\
  \citenamefont {Kourta}}]{debien2016}%
  \BibitemOpen
  \bibfield  {author} {\bibinfo {author} {\bibfnamefont {A.}~\bibnamefont
  {Debien}}, \bibinfo {author} {\bibfnamefont {K.~A. F.~F.}\ \bibnamefont {von
  Krbek}}, \bibinfo {author} {\bibfnamefont {N.}~\bibnamefont {Mazellier}},
  \bibinfo {author} {\bibfnamefont {T.}~\bibnamefont {Duriez}}, \bibinfo
  {author} {\bibfnamefont {L.}~\bibnamefont {Cordier}}, \bibinfo {author}
  {\bibfnamefont {B.~R.}\ \bibnamefont {Noack}}, \bibinfo {author}
  {\bibfnamefont {M.~W.}\ \bibnamefont {Abel}}, \ and\ \bibinfo {author}
  {\bibfnamefont {A.}~\bibnamefont {Kourta}},\ }\bibfield  {title} {\enquote
  {\bibinfo {title} {Closed-loop separation control over a sharp edge ramp
  using genetic programming},}\ }\href@noop {} {\bibfield  {journal} {\bibinfo
  {journal} {Exp. Fluids}\ }\textbf {\bibinfo {volume} {57}},\ \bibinfo {pages}
  {40} (\bibinfo {year} {2016})}\BibitemShut {NoStop}%
\bibitem [{\citenamefont {Schlichting}\ \emph {et~al.}(1968)\citenamefont
  {Schlichting}, \citenamefont {Gersten}, \citenamefont {Krause}, \citenamefont
  {Oertel},\ and\ \citenamefont {Mayes}}]{BLTheory}%
  \BibitemOpen
  \bibfield  {author} {\bibinfo {author} {\bibfnamefont {H.}~\bibnamefont
  {Schlichting}}, \bibinfo {author} {\bibfnamefont {K.}~\bibnamefont
  {Gersten}}, \bibinfo {author} {\bibfnamefont {E.}~\bibnamefont {Krause}},
  \bibinfo {author} {\bibfnamefont {H.}~\bibnamefont {Oertel}}, \ and\ \bibinfo
  {author} {\bibfnamefont {K.}~\bibnamefont {Mayes}},\ }\href@noop {} {\emph
  {\bibinfo {title} {Boundary-layer theory}}},\ \bibinfo {edition} {6th}\ ed.\
  (\bibinfo  {publisher} {Springer},\ \bibinfo {year} {1968})\BibitemShut
  {NoStop}%
\bibitem [{\citenamefont {Guilmineau}\ \emph {et~al.}(2014)\citenamefont
  {Guilmineau}, \citenamefont {Duvigneau},\ and\ \citenamefont
  {Labroquère}}]{guilmineau2014}%
  \BibitemOpen
  \bibfield  {author} {\bibinfo {author} {\bibfnamefont {E.}~\bibnamefont
  {Guilmineau}}, \bibinfo {author} {\bibfnamefont {R.}~\bibnamefont
  {Duvigneau}}, \ and\ \bibinfo {author} {\bibfnamefont {J.}~\bibnamefont
  {Labroquère}},\ }\bibfield  {title} {\enquote {\bibinfo {title}
  {Optimization of jet parameters to control the flow on a ramp},}\ }\href@noop
  {} {\bibfield  {journal} {\bibinfo  {journal} {Comptes Rendus Mécanique:
  Flow separation control}\ }\textbf {\bibinfo {volume} {342}},\ \bibinfo
  {pages} {363--375} (\bibinfo {year} {2014})}\BibitemShut {NoStop}%
\bibitem [{\citenamefont {Roshko}\ and\ \citenamefont
  {Lau}(1965)}]{roshko1965}%
  \BibitemOpen
  \bibfield  {author} {\bibinfo {author} {\bibfnamefont {A.}~\bibnamefont
  {Roshko}}\ and\ \bibinfo {author} {\bibfnamefont {J.~C.}\ \bibnamefont
  {Lau}},\ }\bibfield  {title} {\enquote {\bibinfo {title} {Some observations
  on transition and reattachment of a free shear layer in incompressible
  flow},}\ }in\ \href@noop {} {\emph {\bibinfo {booktitle} {Proceedings of the
  Heat Transfer and Fluid Mechanics Institute}}},\ Vol.~\bibinfo {volume} {18}\
  (\bibinfo  {publisher} {Stanford University Press},\ \bibinfo {year} {1965})\
  pp.\ \bibinfo {pages} {157--167}\BibitemShut {NoStop}%
\end{thebibliography}%

\end{document}